\documentclass[a4paper, authoryear,12pt]{article}
\usepackage{apacite}
\usepackage[pdftex,colorlinks = true]{hyperref}
\usepackage[authoryear,round,longnamesfirst]{natbib}
\usepackage[font=small]{caption}
\usepackage[nolists]{endfloat}
\usepackage{amsmath, natbib, bm, url, setspace, tikz, geometry, microtype, subfig, xcolor, booktabs, todonotes, paralist, dsfont}
\usepackage{geometry}
\hypersetup{citecolor=darkblue,linkcolor=darkblue,urlcolor=darkblue}
\definecolor{darkblue}{rgb}{0,0,.6}
\usepackage[font={small}]{caption}

\usepackage[utf8]{inputenc}
\usepackage{xspace}
\usepackage{xcolor,bm}
\usepackage{graphicx}
\definecolor{Rcolor}{RGB}{150,160,190}
\definecolor{a0}{rgb}{0.0, 0.5, 0.0}
\definecolor{bistre}{rgb}{0.24, 0.17, 0.12}
\definecolor{amethyst}{rgb}{0.6, 0.4, 0.8}
\definecolor{blue-violet}{rgb}{0.54, 0.17, 0.89}
\definecolor{Rcolor}{RGB}{150,160,190}
\definecolor{blush}{rgb}{0.87, 0.36, 0.51}
\definecolor{brightturquoise}{rgb}{0.03, 0.91, 0.87}
\definecolor{purple}{rgb}{0.8, 0.33, 0.0}

\newcommand{\R}{R}
  
\makeatletter

\newsavebox\CBox
\def\textBF#1{\sbox\CBox{#1}\resizebox{\wd\CBox}{\ht\CBox}{\textbf{#1}}}
\graphicspath{{plots/}}

\begin{document}

\begin{center}
\Large Reconciling forecasts of infant mortality rates at national and sub-national levels: Grouped time-series methods
\end{center}
\begin{center}
Han Lin Shang \footnote{Postal address: Research School of Finance, Actuarial Studies and Statistics, Australian National University, Acton ACT 2601, Australia; Telephone number: +61(2) 6125 0535; Fax number: +61(2) 6125 0087; Email address: hanlin.shang@anu.edu.au} \\
Australian National University
\end{center}

\vspace{.3in}

\begin{abstract}

Mortality rates are often disaggregated by different attributes, such as sex, state, education, religion or ethnicity. Forecasting mortality rates at the national and sub-national levels plays an important role in making social policies associated with the national and sub-national levels. However, base forecasts at the sub-national levels may not add up to the forecasts at the national level. To address this issue, we consider the problem of reconciling mortality rate forecasts from the viewpoint of grouped time-series forecasting methods \citep{HAA+11}. A bottom-up method and an optimal combination method are applied to produce point forecasts of infant mortality rates that are aggregated appropriately across the different levels of a hierarchy. We extend these two methods by considering the reconciliation of interval forecasts through a bootstrap procedure. Using the regional infant mortality rates in Australia, we investigate the one-step-ahead to 20-step-ahead point and interval forecast accuracies among the independent and these two grouped time-series forecasting methods. The proposed methods are shown to be useful for reconciling point and interval forecasts of demographic rates at the national and sub-national levels, and would be beneficial for government policy decisions regarding the allocations of current and future resources at both the national and sub-national levels.

\end{abstract}


\newpage

\section{Introduction}

The infant mortality rate is a useful indicator of a country's level of health or development and it is a component of the physical quality of life index. In some societies, sex-specific infant mortality may reveal gender inequalities. For instance, many Asian countries are known to have a preference for sons, which has stimulated research into gender bias, such as in India \citep*{GMH00}, Bangladesh \citep{RD93}, China \citep{CB96}, the Republic of Korea \citep{PC95}, and sub-Saharan Africa \citep*{FK14}. Anomalous female infant mortality is a sign of gender stratification, and as such is in need of detailed investigation by social and medical scientists.

As a part of the United Nations Millennium Development Goals, the infant mortality rate has been widely studied by official statistical agencies worldwide, including the United Nations Statistics Division (\url{http://unstats.un.org/unsd/demographic/products/vitstats}), the United Nations International Children's Emergency Fund (\url{http://www.unicef.org}), the World Health Organization (\url{http://www.who.int/whosis/mort/en}), the World Bank (\url{data.worldbank.org/indicator/SP.DYN.IMRT.IN}), as well as demographic and medical research communities. For example, \cite{FC06} investigated the gender imbalance in infant mortality in a cross-national study consisting of many developing nations, while \cite{DCV+08} studied the rise and fall of excess male infant mortality in a cross-national study consisting of many developed nations. Furthermore, \cite{ABO+06} studied the differences in infant mortality between rural and urban areas in Australia. 

It is not only important to analyze infant mortality by state and examine variations across different states, but also important to analyze the infant mortality rate by sex and examine the hypothesis whether or not the female infant mortality rate will continue to be higher than the male infant mortality rate. With both aggregated and disaggregated historical time series, we aim to model and forecast sex-specific infant mortality rates at national and sub-national levels. When these data are forecast independently without any constraint, we often confront the account balancing problem, where the forecasts at the sub-national level may not add up to the forecasts at the national level. This is known as forecast reconciliation, which has long been studied by \cite{SCM42} and further studied by \cite{Weale88} and \cite{SW09}, in the context of balancing the national economic account. Here, we extend this forecast reconciliation from economics to demography.

The reconciliation methods proposed will not only enhance interpretation of mortality forecasts, but can also improve forecast accuracy as it obeys a group structure. Any improvements in the forecast accuracy of mortality would be beneficial for governments, in particular for determining age of retirements; annuity providers and corporate pension funds for allocating pension benefits at the national and sub-national levels.

To the best of our knowledge, there is little or no work on reconciling forecasts of infant mortality rates at the different levels of a hierarchy, where infant mortality rates can be disaggregated by sex and state. We consider a bottom-up method and an optimal combination method of \cite{HAA+11}, and extend these methods to model rates instead of counts. These methods do not only produce point forecasts for infant mortality rates at the national and sub-national levels, but also the point forecasts at the sub-national level sum up to the forecasts at the national level. As a result, the point forecasts and the original time series both preserve the group structure. The main contribution of this paper is to put forward a bootstrap procedure for constructing prediction intervals for the bottom-up and optimal combination methods, since forecast uncertainty can never be overlooked. 

When we observe multiple time series that are correlated, we often confront the so-called grouped time series. Grouped time series are typically time series organized in a hierarchical structure based on different attributes, such as sex, state, education, religion or ethnicity. For example, \cite{AAH09} disaggregate the Australian tourism demand by states. Tourism demand within each state is then disaggregated into different zones. Tourism demand within each zone is further divided into different regions. In demographic forecasting, the infant mortality rates in Australia can first be disaggregated by sex. Within each sex, mortality rates can then be further disaggregated by the different Australian states. The first example is referred to as a hierarchical time series, in which the order of disaggregation is unique. The second example, which will be studied here, is called a grouped time series. Grouped time series can be thought of as hierarchical time series without a unique hierarchical structure. In other words, the infant mortality rates in Australia can also be first disaggregated by state and then by sex. 

Existing approaches to hierarchical/grouped time-series forecasting in econometrics and statistics usually consider a top-down method, bottom-up method, middle-out method or an optimal combination method. A top-down method predicts the aggregated series at the top level and then disaggregates the forecasts based on historical or forecast proportions \citep[see for example,][]{GS90}. The bottom-up method involves forecasting each of the disaggregated series at the lowest level of the hierarchy and then using simple aggregation to obtain forecasts at the higher levels of the hierarchy \citep[see for example,][]{Kahn98}. In practice, it is common to combine both methods, where forecasts are obtained for each series at an intermediate level of the hierarchy, before aggregating them to the series at the top level and disaggregating them to the series at the bottom level. This method is referred to as the middle-out method. \cite{HAA+11} and \cite{HLW16} proposed an optimal combination method, where base forecasts are obtained independently for all series at all levels of the hierarchy and then a linear regression model is used with an ordinary least squares (OLS) or a generalized least squares (GLS) estimator to optimally combine and reconcile these forecasts. They showed that the revised forecasts do not only add up across the hierarchy, but they are also unbiased and have minimum variance amongst all combined forecasts under some simple assumptions \citep{HAA+11}. 

To the best of our knowledge, these four hierarchical time-series methods are only applicable to counts not rates. Among the four hierarchical time-series forecasting methods, the top-down and middle-out methods are not suitable for analyzing grouped time series because of the non-unique structure of the hierarchy. In Section~\ref{sec:2}, we first revisit a bottom-up and an optimal combination method to produce point forecasts of infant mortality rates, and then propose a bootstrap method to reconcile interval forecasts.  Using the Australian infant mortality rates described in Section~\ref{sec:3}, we investigate the one-step-ahead to 20-step-ahead point and interval forecast accuracies in Sections~\ref{sec:4} and~\ref{sec:5}, respectively. Conclusions are given in Section~\ref{sec:6}, along with some reflections on how the methods developed here might be further extended. In the Appendix, we present some details on maximum entropy bootstrapping.

\section{Some grouped time-series forecasting methods}\label{sec:2}

\subsection{Notation}

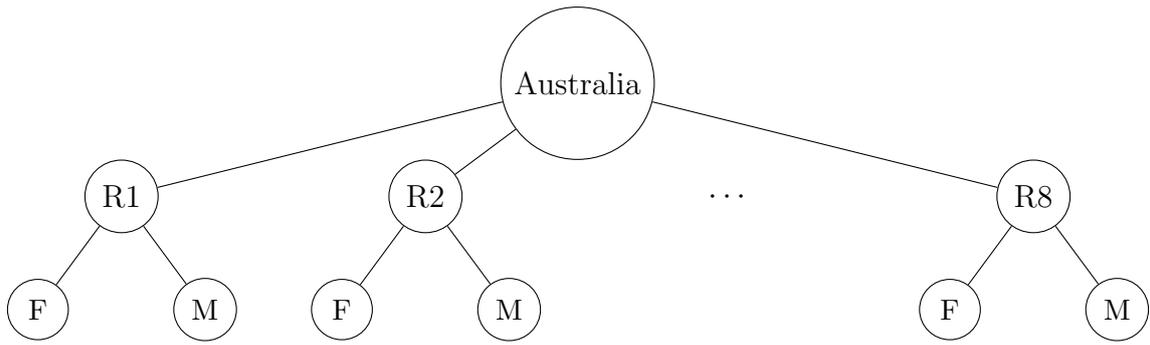
\begin{figure}[!htbp]
\centering\begin{tikzpicture}
\tikzstyle{every node}=[minimum size = 8mm]
\tikzstyle[level distance=10cm] \tikzstyle[sibling distance=40cm]
\tikzstyle{level 3}=[sibling distance=16mm,font=\footnotesize]
\tikzstyle{level 2}=[sibling distance=22mm,font=\small]
\tikzstyle{level 1}=[sibling distance=40mm,font=\normalsize]
\node[circle,draw]{Australia}
   child {node[circle,draw] {R1}
   	     child {node[circle,draw] {F}}
    		child {node[circle,draw] {M}}
}
   child {node[circle,draw] {R2}
   		child {node[circle,draw] {F}}
		child {node[circle,draw] {M}}
		}
  child {node {$\cdots$}edge from parent[draw=none]}
   child {node[circle,draw] {R8}
   		child{node[circle,draw] {F}}
  	child{node[circle,draw] {M}}
 };
\end{tikzpicture}
\caption{A two-level hierarchical tree diagram, with eight regions. In the top level, we have the mortality for Australia; in Level 1, total mortality of Australia can be disaggregated by eight regions; in Level 2, total mortality of each region can be disaggregated by sex within each region.}\label{fig:1}
\end{figure}

For ease of explanation, we will introduce the notation using the Australian data example (see Section~\ref{sec:31} for more details). The generalization to other contexts should be apparent. The Australian data follow a multilevel geographical hierarchy coupled with a sex grouping variable. The geographical hierarchy is shown in Fig.~\ref{fig:1}, where Australia is split into eight regions.

Let $\bm{C}_t = \left[C_t, \bm{C}^{\top}_{1,t}, \dots, \bm{C}^{\top}_{K,t}\right]^{\top}$, where $C_t$ is the total of all series at time $t=1,2,\dots,n$, $\bm{C}_{k,t}$ represents the vector of all observations at level $k$ at time $t$ and $^{\top}$ symbolizes the matrix transpose. As shown in Fig.~\ref{fig:1}, counts at higher levels can be obtained by summing the series below.
\begin{align*}
C_{t}& = C_{\text{R1},t} + C_{\text{R2},t} + \cdots + C_{\text{R8},t},\\
C_{\text{R1},t} &= C_{\text{R1}*\text{F},t}+C_{\text{R1}*\text{M},t}.
\end{align*}
Alternatively, we can also express the hierarchy using a matrix notation \citep[see][]{AAH09}. Note that
\begin{equation*}
\bm{C}_t = \bm{S}\times\bm{C}_{K,t},
\end{equation*}
where $\bm{S}$ is a ``summing" matrix of order $m\times m_K$, $m$ represents the total number of series ($1+2+8+16=27$ for the hierarchy in Fig.~\ref{fig:1}) and $m_K$ represents the total number of bottom-level series. The summing matrix $\bm{S}$, which delineates how the bottom-level series are aggregated, is consistent with the group structure. For modeling mortality \textit{counts}, we can express the hierarchy in Fig.~\ref{fig:1} as
\begin{footnotesize}
\arraycolsep=0.1cm
\[
\underbrace{ \left[
\begin{array}{l}
C_{\text{T},t} \\
C_{\textcolor{red}{\text{F},t}} \\
C_{\textcolor{red}{\text{M},t}} \\
C_{\textcolor{a0}{\text{R1*T},t}} \\
C_{\textcolor{a0}{\text{R2*T},t}} \\
\vdots \\
C_{\textcolor{a0}{\text{R8*T},t}} \\
C_{\textcolor{purple}{\text{R1*F},t}} \\
C_{\textcolor{purple}{\text{R1*M},t}} \\
C_{\textcolor{purple}{\text{R2*F},t}} \\
C_{\textcolor{purple}{\text{R2*M},t}} \\
\vdots \\
C_{\textcolor{purple}{\text{R8*F},t}} \\
C_{\textcolor{purple}{\text{R8*M},t}} \\
 \end{array}
\right]}_{\bm{C}_t} =
\underbrace{\left[
\begin{array}{ccccccccccc}
1 & 1 & 1 & 1 & 1 & 1 & \cdots & 1 & 1 \\
\textcolor{red}{1} & \textcolor{red}{0} & \textcolor{red}{1} & \textcolor{red}{0} & \textcolor{red}{1} & \textcolor{red}{0} & \cdots & \textcolor{red}{1} & \textcolor{red}{0} \\
\textcolor{red}{0} & \textcolor{red}{1}  & \textcolor{red}{0} & \textcolor{red}{1} & \textcolor{red}{0} & \textcolor{red}{1} & \cdots & \textcolor{red}{0} & \textcolor{red}{1} \\
\textcolor{a0}{1} & \textcolor{a0}{1} & \textcolor{a0}{0} & \textcolor{a0}{0} & \textcolor{a0}{0} & \textcolor{a0}{0} & \cdots  & \textcolor{a0}{0} & \textcolor{a0}{0} \\
\textcolor{a0}{0} & \textcolor{a0}{0} & \textcolor{a0}{1} & \textcolor{a0}{1} & \textcolor{a0}{0} & \textcolor{a0}{0} & \cdots & \textcolor{a0}{0} & \textcolor{a0}{0} \\
\vdots & \vdots & \vdots & \vdots & \vdots & \vdots & \cdots & \vdots & \vdots \\
\textcolor{a0}{0} & \textcolor{a0}{0} & \textcolor{a0}{0} & \textcolor{a0}{0} & \textcolor{a0}{0} & \textcolor{a0}{0} & \cdots & \textcolor{a0}{1} & \textcolor{a0}{1} \\
\textcolor{purple}{1} & \textcolor{purple}{0} & \textcolor{purple}{0} & \textcolor{purple}{0} & \textcolor{purple}{0} & \textcolor{purple}{0} & \cdots & \textcolor{purple}{0} & \textcolor{purple}{0} \\
\textcolor{purple}{0} & \textcolor{purple}{1} & \textcolor{purple}{0} & \textcolor{purple}{0} & \textcolor{purple}{0} & \textcolor{purple}{0} & \cdots & \textcolor{purple}{0} & \textcolor{purple}{0} \\
\textcolor{purple}{0} & \textcolor{purple}{0} & \textcolor{purple}{1} & \textcolor{purple}{0} & \textcolor{purple}{0} & \textcolor{purple}{0} & \cdots & \textcolor{purple}{0} & \textcolor{purple}{0} \\
\textcolor{purple}{0} & \textcolor{purple}{0} & \textcolor{purple}{0} & \textcolor{purple}{1} & \textcolor{purple}{0} & \textcolor{purple}{0} & \cdots & \textcolor{purple}{0} & \textcolor{purple}{0} \\
\vdots & \vdots & \vdots & \vdots & \vdots & \vdots & \cdots & \vdots & \vdots  \\
\textcolor{purple}{0} & \textcolor{purple}{0} & \textcolor{purple}{0} & \textcolor{purple}{0} & \textcolor{purple}{0} & \textcolor{purple}{0} & \cdots  & \textcolor{purple}{1} & \textcolor{purple}{0}\\
\textcolor{purple}{0} & \textcolor{purple}{0} & \textcolor{purple}{0} & \textcolor{purple}{0} & \textcolor{purple}{0} & \textcolor{purple}{0} & \cdots & \textcolor{purple}{0} & \textcolor{purple}{1} \\
\end{array}
\right]}_{\bm{S}_t}
\underbrace{\left[
\begin{array}{l}
C_{\text{R1*F},t} \\
C_{\text{R1*M},t} \\
C_{\text{R2*F},t} \\
C_{\text{R2*M},t} \\
\vdots \\
C_{\text{R8*F},t} \\
C_{\text{R8*M},t} \\
\end{array}
\right]}_{\bm{C}_{K,t}}.
\]
\end{footnotesize}

For modeling mortality \textit{rates}, we can express the hierarchy in Fig.~\ref{fig:1} as
\begin{footnotesize}
\arraycolsep=0.1cm
\[
\underbrace{ \left[
\begin{array}{l}
R_{\text{T},t} \\
R_{\textcolor{red}{\text{F},t}} \\
R_{\textcolor{red}{\text{M},t}} \\
R_{\textcolor{a0}{\text{R1*T},t}} \\
R_{\textcolor{a0}{\text{R2*T},t}} \\
\vdots \\
R_{\textcolor{a0}{\text{R8*T},t}} \\
R_{\textcolor{purple}{\text{R1*F},t}} \\
R_{\textcolor{purple}{\text{R1*M},t}} \\
R_{\textcolor{purple}{\text{R2*F},t}} \\
R_{\textcolor{purple}{\text{R2*M},t}} \\
\vdots \\
R_{\textcolor{purple}{\text{R8*F},t}} \\
R_{\textcolor{purple}{\text{R8*M},t}} \\
 \end{array}
\right]}_{\bm{R}_t} =
\underbrace{\left[
\begin{array}{ccccccccccc}
\frac{E_{\text{R1*F},t}}{E_{\text{T},t}} & \frac{E_{\text{R1*M},t}}{E_{\text{T},t}} & \frac{E_{\text{R2*F},t}}{E_{\text{T},t}} & \frac{E_{\text{R2*M},t}}{E_{\text{T},t}}  & \frac{E_{\text{R3*F},t}}{E_{\text{T},t}} & \frac{E_{\text{R3*M},t}}{E_{\text{T},t}} & \cdots & \frac{E_{\text{R8*F},t}}{E_{\text{T},t}} & \frac{E_{\text{R8*M},t}}{E_{\text{T},t}} \\
\textcolor{red}{\frac{E_{\text{R1*F},t}}{E_{\text{F},t}}} & \textcolor{red}{0} & \textcolor{red}{\frac{E_{\text{R2*F},t}}{E_{\text{F},t}}} & \textcolor{red}{0} & \textcolor{red}{\frac{E_{\text{R3*F},t}}{E_{\text{F},t}}} & \textcolor{red}{0} & \cdots & \textcolor{red}{\frac{E_{\text{R8*F},t}}{E_{\text{F},t}}} & \textcolor{red}{0} \\
\textcolor{red}{0} & \textcolor{red}{\frac{E_{\text{R1*M},t}}{E_{\text{M},t}}}  & \textcolor{red}{0} & \textcolor{red}{\frac{E_{\text{R2*M},t}}{E_{\text{M},t}}} & \textcolor{red}{0} & \textcolor{red}{\frac{E_{\text{R3*M},t}}{E_{\text{M},t}}} & \cdots & \textcolor{red}{0} & \textcolor{red}{\frac{E_{\text{R8*M},t}}{E_{\text{M},t}}} \\
\textcolor{a0}{\frac{E_{\text{R1*F},t}}{E_{\text{R1*T},t}}} & \textcolor{a0}{\frac{E_{\text{R1*M},t}}{E_{\text{R1*T},t}}} & \textcolor{a0}{0} & \textcolor{a0}{0} & \textcolor{a0}{0} & \textcolor{a0}{0} & \cdots  & \textcolor{a0}{0} & \textcolor{a0}{0} \\
\textcolor{a0}{0} & \textcolor{a0}{0} & \textcolor{a0}{\frac{E_{\text{R2*F},t}}{E_{\text{R2*T},t}}} & \textcolor{a0}{\frac{E_{\text{R2*M},t}}{E_{\text{R2*T},t}}} & \textcolor{a0}{0} & \textcolor{a0}{0} & \cdots & \textcolor{a0}{0} & \textcolor{a0}{0} \\
\vdots & \vdots & \vdots & \vdots & \vdots & \vdots & \cdots & \vdots & \vdots \\
\textcolor{a0}{0} & \textcolor{a0}{0} & \textcolor{a0}{0} & \textcolor{a0}{0} & \textcolor{a0}{0} & \textcolor{a0}{0} & \cdots & \textcolor{a0}{\frac{E_{\text{R8*F},t}}{E_{\text{R8*T},t}}} & \textcolor{a0}{\frac{E_{\text{R8*M},t}}{E_{\text{R8*T},t}}} \\
\textcolor{purple}{1} & \textcolor{purple}{0} & \textcolor{purple}{0} & \textcolor{purple}{0} & \textcolor{purple}{0} & \textcolor{purple}{0} & \cdots & \textcolor{purple}{0} & \textcolor{purple}{0} \\
\textcolor{purple}{0} & \textcolor{purple}{1} & \textcolor{purple}{0} & \textcolor{purple}{0} & \textcolor{purple}{0} & \textcolor{purple}{0} & \cdots & \textcolor{purple}{0} & \textcolor{purple}{0} \\
\textcolor{purple}{0} & \textcolor{purple}{0} & \textcolor{purple}{1} & \textcolor{purple}{0} & \textcolor{purple}{0} & \textcolor{purple}{0} & \cdots & \textcolor{purple}{0} & \textcolor{purple}{0} \\
\textcolor{purple}{0} & \textcolor{purple}{0} & \textcolor{purple}{0} & \textcolor{purple}{1} & \textcolor{purple}{0} & \textcolor{purple}{0} & \cdots & \textcolor{purple}{0} & \textcolor{purple}{0} \\
\vdots & \vdots & \vdots & \vdots & \vdots & \vdots & \cdots & \vdots & \vdots  \\
\textcolor{purple}{0} & \textcolor{purple}{0} & \textcolor{purple}{0} & \textcolor{purple}{0} & \textcolor{purple}{0} & \textcolor{purple}{0} & \cdots  & \textcolor{purple}{1} & \textcolor{purple}{0}\\
\textcolor{purple}{0} & \textcolor{purple}{0} & \textcolor{purple}{0} & \textcolor{purple}{0} & \textcolor{purple}{0} & \textcolor{purple}{0} & \cdots & \textcolor{purple}{0} & \textcolor{purple}{1} \\
\end{array}
\right]}_{\bm{S}_t}
\underbrace{\left[
\begin{array}{l}
R_{\text{R1*F},t} \\
R_{\text{R1*M},t} \\
R_{\text{R2*F},t} \\
R_{\text{R2*M},t} \\
\vdots \\
R_{\text{R8*F},t} \\
R_{\text{R8*M},t} \\
\end{array}
\right]}_{\bm{R}_{K,t}},
\]
\end{footnotesize}
\hspace{-.15in} where $\textit{E}_{\text{R1}* \text{F},t}/\textit{E}_{\text{T},t}$ represents the ratio between the exposure-to-risk for female series in Region 1 and the exposure-to-risk for total series in entire Australia at time $t$, and $R_{\text{R1}*\text{F},t} = \textit{D}_{\text{R1}*\text{F},t}/\textit{E}_{\text{R1}*\text{F},t}$ represents the mortality rate given by the ratio between the number of deaths and exposure-to-risk for female series in region 1 at time $t$, for $t=1,\dots,n$. 

Based on the information available up to and including time $n$, we are interested in computing forecasts for each series at each level, giving $m$ base forecasts for the forecasting period $n+h,\dots,n+w$, where $h$ represents the forecast horizon and $w\geq h$ represents the last year of the forecasting period. We denote 
\begin{itemize}
\item $\widehat{R}_{\text{T},n+h}$ as the $h$-step-ahead base forecast of Series Total in the forecasting period, 
\item $\widehat{R}_{\text{R1},n+h}$ as the $h$-step-ahead forecast of the series Region 1, and 
\item $\widehat{R}_{\text{R1}*\text{F},n+h}$ as the $h$-step-ahead forecast of the female series in Region 1. 
\end{itemize}
These base forecasts can be obtained for each series in the hierarchy using a suitable forecasting method, such as the automatic autoregressive integrated moving average (ARIMA) \citep{HK08} implemented here. They are then combined in such ways to produce final forecasts for the whole hierarchy that aggregate in a manner which is consistent with the structure of the hierarchy. We refer to these as revised forecasts and denote them as $\overline{R}_{\text{T},n+h}$ and $\overline{\bm{R}}_{k,n+h}$ for level $k=1,\dots,K$.

In the following sections, we describe two ways of combining the base forecasts in order to obtain revised forecasts. These two methods were originally proposed for modeling counts, here we extend these methods for modeling rates.

\subsection{Bottom-up method}

One of the commonly used methods for hierarchical/grouped time-series forecasting is the bottom-up method \citep[e.g.,][]{Kinney71, DM92, ZT00}. This method involves first generating base forecasts for each series at the bottom level of the hierarchy and then aggregating these upwards to produce revised forecasts for the whole hierarchy. As an example, let us consider the hierarchy of Fig.~\ref{fig:1}. We first generate $h$-step-ahead base forecasts for the bottom-level series, namely $\widehat{R}_{\text{R}_1*\text{F},n+h},\widehat{R}_{\text{R}_1*\text{M},n+h}, \widehat{R}_{\text{R}_2*\text{F},n+h},\widehat{R}_{\text{R}_2*\text{M},n+h}, \cdots,$ $\widehat{R}_{\text{R}_8*\text{F},n+h},\widehat{R}_{\text{R}_8*\text{M},n+h}$. Aggregating these up the hierarchy, we get $h$-step-ahead forecasts for the rest of series, as stated below.
\begin{itemize}
\item $\overline{R}_{\text{F},n+h} = \frac{E_{\text{R1*F},n+h}}{E_{\text{T},n+h}} \times \widehat{R}_{\text{R1*F},n+h}+\frac{E_{\text{R2*F},n+h}}{E_{\text{T},n+h}}\times\widehat{R}_{\text{R2*F},n+h}+\cdots + \frac{E_{\text{R8*F},n+h}}{E_{\text{T},n+h}}\times\widehat{R}_{\text{R8*F},n+h},$  
\item $\overline{R}_{\text{M},n+h} = \frac{E_{\text{R1*M},n+h}}{E_{\text{T},n+h}} \times \widehat{R}_{\text{R1*M},n+h}+\frac{E_{\text{R2*M},n+h}}{E_{\text{T},n+h}}\times\widehat{R}_{\text{R2*M},n+h}+\cdots + \frac{E_{\text{R8*M},n+h}}{E_{\text{T},n+h}}\times\widehat{R}_{\text{R8*M},n+h}, \text{and}$
\item $\overline{R}_{n+h} = \frac{E_{\text{F},n+h}}{E_{\text{T},n+h}}\times\overline{R}_{\text{F},n+h}+\frac{E_{\text{M},n+h}}{E_{\text{T},n+h}}\times\overline{R}_{\text{M},n+h}$,
\end{itemize}
where $\overline{R}_{\text{F},n+h}$ and $\overline{R}_{\text{M},n+h}$ represent reconciled forecasts. The revised forecasts for the bottom-level series are the same as the base forecasts in the bottom-up method (i.e., $\overline{R}_{\text{R1*F},n+h}=\widehat{R}_{\text{R1*F},n+h}$).

The bottom-up method can also be expressed by the summing matrix and we write
\begin{equation*}
\overline{\bm{R}}_{n+h} = \bm{S} \times \widehat{\bm{R}}_{K,n+h},
\end{equation*}
where $\overline{\bm{R}}_{n+h}=\left[\overline{R}_{n+h},\overline{\bm{R}}_{1,n+h}^{\top},\dots,\overline{\bm{R}}_{K,n+h}^{\top}\right]^{\top}$ represents the revised forecasts for the whole hierarchy and $\widehat{\bm{R}}_{K,n+h}$ represents the bottom-level forecasts.

The bottom-up method has an agreeable feature in that no information is lost due to aggregation, and it performs well when the signal-to-noise ratio is strong at the bottom-level series. On the other hand, it may lead to inaccurate forecasts of the top-level series, when there are many missing or noisy data at the bottom level \citep[see for example,][]{SW79, STM88}.

\subsection{Optimal combination}

This method involves first producing base forecasts independently for each time series at each level of a hierarchy. As these base forecasts are independently generated, they will not be `aggregate consistent' (i.e., they will not sum appropriately according to the group structure). The optimal combination method optimally combines the base forecasts through linear regression by generating a set of revised forecasts that are as close as possible to the base forecasts but that also aggregate consistently within the group. The essence is derived from the representation of $h$-step-ahead base forecasts for the entire hierarchy by linear regression. That is,
\begin{equation*}
\widehat{\bm{R}}_{n+h} = \bm{S}\times \bm{\beta}_{n+h} + \bm{\varepsilon}_{n+h},
\end{equation*}
where $\widehat{\bm{R}}_{n+h}$ is a vector of the $h$-step-ahead base forecasts for the entire hierarchy, stacked in the same hierarchical order as for original data matrix $\bm{R}_t$ for $t=1,\dots,n$; $\bm{\beta}_{n+h}=\text{E}[\bm{R}_{K,n+h}|\bm{R}_1,\dots,\bm{R}_n]$ is the unknown mean of the base forecasts of the bottom level $K$; and $\bm{\varepsilon}_{n+h}$ represents the estimation errors in the regression, which has zero mean and unknown covariance matrix $\bm{\Sigma}_h$.

Given the base forecasts approximately satisfy the group aggregation structure (which should occur for any reasonable set of forecasts), the errors approximately satisfy the same aggregation structure as the data. That is,
\begin{equation}
\bm{\varepsilon}_{n+h} \approx \bm{S}\times\bm{\varepsilon}_{K,n+h}, \label{eq:OLS}
\end{equation}
where $\bm{\varepsilon}_{K,n+h}$ represents the forecast errors in the bottom level. Under this assumption, \citet[][Theorem 1]{HAA+11} show that the best linear unbiased estimator for $\bm{\beta}_{n+h}$ is
\begin{equation*}
\widehat{\bm{\beta}}_{n+h} = \left(\bm{S}^{\top}\bm{\Sigma}_h^{+}\bm{S}\right)^{-1}\bm{S}^{\top}\bm{\Sigma}_h^+\widehat{\bm{R}}_{n+h},
\end{equation*}
where $\bm{\Sigma}_h^+$ denotes the Moore-Penrose generalized inverse of $\bm{\Sigma}_h$. The revised forecasts are then given by
\begin{equation*}
\overline{\bm{R}}_{n+h} = \bm{S}\times\widehat{\bm{\beta}}_{n+h}.
\end{equation*}
The revised forecasts are unbiased, since $\bm{S}\left(\bm{S}^{\top}\bm{\Sigma}_h^{+}\bm{S}\right)^{-1}\bm{S}^{\top}\bm{\Sigma}_h^+=\bm{I}_m$ where $\bm{I}_m$ denotes $(m\times m)$ identity matrix and $m$ represents the total number of series; the revised forecasts have minimum variances $\text{Var}[\overline{\bm{R}}_{n+h}|\bm{R}_1,\dots,\bm{R}_n] = \bm{S}\left(\bm{S}^{\top}\bm{\Sigma}_h^+\bm{S}\right)^{-1}\bm{S}^{\top}$. 

Under the assumption given in Eq.~\eqref{eq:OLS}, the estimation problem reduces from GLS to OLS, thus it is ideal for handling large-dimensional covariance structures. Even if the aggregation errors do not satisfy this assumption, the OLS solution will still be a consistent way of reconciling the base forecasts \citep{HLW16}. On the other hand, it is possible that assumption~\eqref{eq:OLS} becomes less and less adequate, in particular for a longer and longer forecast horizon. 

\cite{HLW16} proposed a GLS estimator, where the elements of $\bm{\Sigma}_h^+$ are set to the inverse of the variances of the base forecasts, $\text{Var}(y_{n+1}-\widehat{y}_{n+1|n})$. Note that we use the one-step-ahead forecast variances, not the $h$-step-ahead forecast variances. This is because the one-step-ahead forecast variances are readily available as the residual variances for each of the base forecasting models. We assume that these are approximately proportional to the $h$-step-ahead forecast variances, which is true for almost all standard time series forecasting models \citep[see e.g.,][]{HKO+08}. 

\subsection{Univariate time-series forecasting method}

For each series given in Table~\ref{tab:3}, we consider a univariate time-series forecasting method, namely the automatic ARIMA method. This univariate time-series forecasting method is able to model non-stationary time series containing a stochastic trend component. As the yearly mortality data do not contain seasonality, the ARIMA has the general form:
\begin{equation*}
\left(1-\phi_1 B - \cdots - \phi_pB^p\right)\left(1-B\right)^dx_t = \gamma + \left(1+\theta_1B + \cdots + \theta_qB^q\right)w_t,
\end{equation*}
where $\gamma$ represents the intercept, $\left(\phi_1,\cdots,\phi_p\right)$ represent the coefficients associated with the autoregressive component, $\left(\theta_1,\cdots,\theta_q\right)$ represent the coefficients associated with the moving average component, $B$ denotes the backshift operator, and $d$ denotes the order of integration. We use the automatic algorithm of \cite{HK08} to choose the optimal orders of autoregressive $p$, moving average $q$ and difference order $d$. $d$ is selected based on successive Kwiatkowski-Phillips-Schmidt-Shin (KPSS) unit-root test \citep{KPSS92}. KPSS tests are used for testing the null hypothesis that an observable time series is stationary around a deterministic trend. We first test the original time series for a unit root; if the test result is significant, then we test the differenced time series for a unit root. The procedure continues until we obtain our first insignificant result. Having determined $d$, the orders of $p$ and $q$ are selected based on the optimal Akaike information criterion (AIC) with a correction for small sample sizes \citep{Akaike74}. Having identified the optimal ARIMA model, maximum likelihood method can then be used to estimate the parameters. 

Note that instead of ARIMA, other univariate time-series forecasting methods (such as exponential smoothing models \citep{HKO+08}) or multivariate time-series forecasting methods (such as vector autoregressive models \citep{Lutkepohl06}) can be used. However, as the effort in comparing forecast accuracy obtained from these models might distract too much from our emphasis on forecast reconciliation, we do not address these other methods in this paper. Instead, we save discussion of these other models for future research.

\subsection{Prediction interval construction}\label{sec:2.4}

To construct a prediction interval, we consider a combination of the maximum entropy bootstrap proposed by \cite{Vinod04} and a parametric bootstrap. The parametric bootstrap captures the forecast uncertainty in the underlying time-series extrapolation models. In contrast to nonparametric bootstrap, the parametric bootstrap method is comparably fast to compute when the group structure contains many sub-national series; and it also enjoys an optimal convergence rate when the underlying parametric model assumptions are satisfied. These assumptions include: the order of ARIMA model is selected correctly and the parameters are estimated correctly. 

\cite{Kilian01} pointed out that the adverse consequences of bootstrapping an over-parameterized model are much less severe than those of bootstrapping an under-parameterized model, and suggested the optimal order selection be based on the AIC rather than the Bayesian Information Criterion. By using the AIC, the parametric bootstrap algorithm conditions on the lag order estimates from the original time series as though they were the true lag orders. In other words, the parametric bootstrapping ignores the sampling uncertainty associated with the lag order estimates and may lead to erroneous inferences \citep[see][Chapter 8 for examples when parametric bootstrap is invalid]{CL11}. As a possible remedy, the maximum entropy bootstrap generates a set of bootstrap samples from the original time series. From bootstrapped samples, the optimal orders selected are allowed to be different and do not necessarily condition on the lag order estimates from the original time series. Instead, the maximum entropy bootstrap re-estimates the lag order in each bootstrap sample. 

The maximum entropy bootstrap possesses several advantages:
\begin{enumerate}[(1)]
\item stationarity is not required;
\item the bootstrap technique computes the ranks of a time series; since the ranks of observations are invariant under a large class of monotone transformations, this invariance property yields robustness of rank-based statistics against outliers and other distributional departures;
\item bootstrap samples satisfy the ergodic theorem, central limit theorem and mean-preserving constraint;
\item it is suitable for panel time series, where the cross covariance of the original time series is reasonably well preserved.
\end{enumerate}
The methodology and an algorithm of the maximum entropy bootstrap are described in \cite{VL09}. In the Appendix, we have briefly outlined the maximum entropy bootstrap algorithm. Computationally, the \verb meboot.pdata.frame \ function in the \textit{meboot} package \citep{VL09} in \R{} language \citep{Team13} was utilized for producing bootstrap samples for all the time series at different levels of a hierarchy. These bootstrap samples are capable of mimicking the correlation within and between original multiple time series. 

For each bootstrapped time series, we then fitted an optimal ARIMA model (see Section~\ref{sec:4}). Assuming the fitted ARIMA model is correct, future sample paths of mortality rates and exposure-to-risk are separately simulated. As for the two grouped time-series forecasting methods, those simulated forecasts are reconciled through the summing matrix. With a set of the bootstrapped forecasts, we can assess the forecast uncertainty by constructing the prediction intervals using corresponding $\alpha/2$ and $1-\alpha/2$ quantiles, at a specified nominal coverage probability denoted by $1-\alpha$. By averaging the prediction intervals over all bootstrapped samples, we obtained an averaged prediction interval. For a reasonably large level of significance $\alpha$, such as $\alpha=0.2$, averaging prediction intervals works well as we estimate the center distribution of the quantiles.

\section{Data sets}\label{sec:3}

\subsection{Australian infant mortality rates}\label{sec:31}

We apply the bottom-up and optimal combination methods to model and forecast infant mortality rates across the different sexes and states in Australia. For each series, we have yearly observations on the infant mortality rates from 1933 to 2003. This data set was obtained from the Australian Social Science Data Archive (\url{http://www.assda.edu.au/}) and is also publicly available in the \textit{addb} package \citep{Hyndman10} in the \R{} language.

The structure of the hierarchy is displayed in Table~\ref{tab:3}. At the top level, we have the total infant mortality rates for Australia. At Level 1, we can split these total rates by sex, although we note the possibility of splitting the total rates by region. At Level 2, the total rates are disaggregated by eight different regions of Australia: New South Wales (NSW), Victoria (VIC), Queensland (QLD), South Australia (SA), Western Australia (WA), Tasmania (TAS), the Australian Capital Territory and the Overseas Territories (ACTOT), and the Northern Territory (NT). At the bottom level, the total rates are disaggregated by different regions of Australia for each sex. This gives 16 series at the bottom level and 27 series in total.

\begin{table}[!ht]
\tabcolsep 0.38in
\centering
\begin{tabular}{@{}llr@{}}\toprule
Level & & Number of series  \\\hline
Australia & & 1	 \\
Sex  & & 2 \\
State       & & 8	\\
Sex  $\times$ State & & 16 \\\hline
Total & & 27 \\\bottomrule
\end{tabular}
\caption{Hierarchy of Australian infant mortality rates.}\label{tab:3}
\end{table}

\begin{figure}[!ht]
\centering
\includegraphics[width=\textwidth]{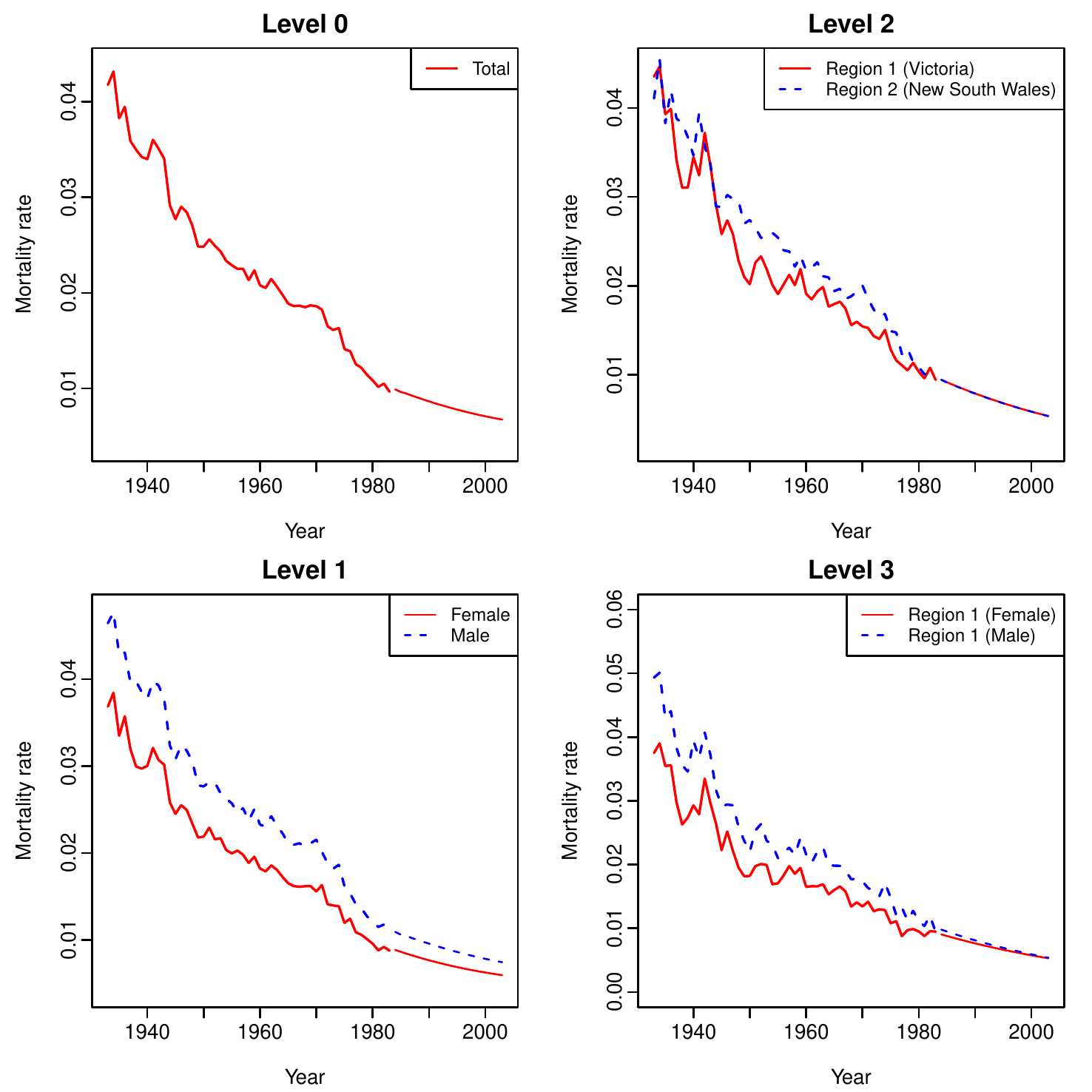}
\caption{Infant mortality rates can be disaggregated by sex in Level 1, region in Level 2, and sex and region in Level 3. For clarity of presentation, we plot only two regions in Level 2, and two sexes of region 1 in Level 3. Based on the data from 1933 to 1983, the bottom-up method is used to produce 20-steps-ahead forecasts from 1984 to 2003 across different levels of the hierarchy. The thicker line(s) represent(s) the historical data, while the thinner line(s) represent(s) the forecasts.}\label{fig:3}
\end{figure}

Figure~\ref{fig:3} shows a few selected series of the infant mortality rates disaggregated by sex, state, and sex and state. As an illustration, based on the data from 1933 to 1983, we apply the bottom-up method to forecast infant mortality rates from 1984 to 2003. The forecasts indicate a continuing decline in infant mortality rates, due largely to improved health services. Moreover, the male infant mortality rates are slightly higher than the female infant mortality rates in Australia. This confirms the early findings of \cite{DCV+08} and \cite{Pongou13}, and it can be explained by environmental causes and also by sex differences in genetic structure and biological makeup, with boys being biologically weaker and more susceptible to diseases and premature death. 

\section{Results of the point forecasts}\label{sec:4}

\subsection{Point forecast evaluation}

A rolling window analysis of a time-series model is commonly used to assess model and parameter stabilities over time. It assesses the constancy of a model's parameter by computing parameter estimates and their forecasts over a rolling window of a fixed size through the sample \citep[see][Chapter 9 for details]{ZW06}. Using the first 51 observations from 1933 to 1983 in the Australian infant mortality rates, we produce one to 20-step-ahead point forecasts. Through a rolling windows approach, we re-estimate the parameters in the univariate time-series forecasting models using the first 52 observations from 1933 to 1984. Forecasts from the estimated models are then produced for one to 19-step-ahead. We iterate this process by increasing the sample size by one year until reaching the end of the data period in 2003. This process produces 20 one-step-ahead forecasts, 19 two-step-ahead forecasts, 18 three-step-ahead forecasts, etc., and one 20-step-ahead forecast. We compare these forecasts with the holdout samples to determine the out-of-sample point forecast accuracy. 

To evaluate the point forecast accuracy, we use the mean absolute forecast error (MAFE) and root mean squared forecast error (RMSFE), which are the absolute and squared percentage errors averaged across years in the forecasting period. As two measures of accuracy, the MAFE and RMSFE show the average difference between estimated and actual populations, regardless of whether the individual estimates were too high or too low. As a measure of bias, the mean forecast error (MFE) shows the average of errors. For each series $j$, they can be defined as
\begin{align*}
 \text{MFE}_{j}(h) &=  \frac{1}{(21-h)}\sum^{n+(20-h)}_{\omega=n}(R_{\omega+h,j}-\widehat{R}_{\omega+h,j}), \\
 \text{MAFE}_{j}(h) &=  \frac{1}{(21-h)}\sum^{n+(20-h)}_{\omega=n}\Big|R_{\omega+h,j}-\widehat{R}_{\omega+h,j}\Big|,\quad \text{and} \\
  \text{RMSFE}_{j}(h) &=  \sqrt{\frac{1}{(21-h)}\sum^{n+(20-h)}_{\omega=n}\left(R_{\omega+h,j}-\widehat{R}_{\omega+h,j}\right)^2}, 
\end{align*}
where $n$ denotes the sample size used for the fitting period for $h=1,2,\dots,20$. By averaging MFE$_{j}(h)$, MAFE$_{j}(h)$ and RMSFE$_j(h)$ across the number of series within each level of a hierarchy, we obtain an overall assessment of the bias and point forecast accuracy for each level and horizon within a hierarchy, denoted by MFE($h$), MAFE$(h)$ and RMSFE$(h)$. They are defined as
\begin{align*}
\text{MFE}(h) & = \frac{1}{m_k}\sum^{m_k}_{j=1}\text{MFE}_j(h), \\
\text{MAFE}(h) &= \frac{1}{m_k}\sum^{m_k}_{j=1}\text{MAFE}_j(h), \quad \text{and} \\
\text{RMSFE}(h) &= \frac{1}{m_k}\sum^{m_k}_{j=1}\text{RMSFE}_j(h), 
\end{align*}
where $m_k$ denotes the number of series at the $k^{\text{th}}$ level of the hierarchy, for $k=1,\dots,K$.

\subsection{Point forecast accuracy of Australian infant mortality rates}

In Tables~\ref{tab:AUS_MFE}, ~\ref{tab:AUS_MAFE} and~\ref{tab:AUS_RMSFE}, we present the MFE$(h)$, MAFE$(h)$ and RMSFE$(h)$ for each level of the hierarchy using the bottom-up and optimal combination methods, and a base forecasting method (i.e., without reconciling forecasts). For ease of comparison, we highlight in bold the method that performs the best for each level of the hierarchy and each forecast horizon, defined as the method with the smallest MFE($h$), MAFE$(h)$ and RMSFE$(h)$.

\begin{table}[!htbp]
\centering
\tabcolsep 0.108in
\begin{small}
\begin{tabular}{@{}lrrrr|rrrr@{}}
\toprule \\[-3.45ex] 
& Total & Sex & Region & Sex$\times$ Region & Total & Sex & Region & Sex$\times$ Region \\
$h$  & \multicolumn{4}{c|}{Base} & \multicolumn{4}{c}{Bottom-up}  \\
  \hline
  1 & -0.004 & -0.005 & -0.037 & -0.065 & -0.013 & -0.013 & -0.065 & -0.065 \\
  2 & -0.005 & -0.006 & -0.052 & -0.081 & -0.016 & -0.016 & -0.081 & -0.081 \\
  3 & -0.015 & -0.016 & -0.074 & -0.106 & -0.030 & -0.030 & -0.106 & -0.106 \\
  4 & -0.021 & -0.022 & -0.095 & -0.126 & -0.039 & -0.039 & -0.125 & -0.126 \\
  5 & -0.028 & -0.028 & -0.114 & -0.147 & -0.050 & -0.049 & -0.147 & -0.147 \\
  6 & -0.039 & -0.040 & -0.140 & -0.175 & -0.065 & -0.065 & -0.175 & -0.175 \\
  7 & -0.049 & -0.050 & -0.164 & -0.199 & -0.079 & -0.078 & -0.199 & -0.199 \\
  8 & -0.063 & -0.064 & -0.191 & -0.229 & -0.097 & -0.097 & -0.230 & -0.229 \\
  9 & -0.072 & -0.073 & -0.215 & -0.255 & -0.110 & -0.110 & -0.256 & -0.255 \\
  10 & -0.085 & -0.086 & -0.242 & -0.283 & -0.127 & -0.127 & -0.284 & -0.283 \\
  11 & -0.090 & -0.091 & -0.259 & -0.304 & -0.136 & -0.136 & -0.305 & -0.304 \\
  12 & -0.090 & -0.091 & -0.270 & -0.319 & -0.141 & -0.141 & -0.321 & -0.319 \\
  13 & -0.089 & -0.090 & -0.284 & -0.338 & -0.147 & -0.146 & -0.340 & -0.338 \\
  14 & -0.085 & -0.087 & -0.296 & -0.356 & -0.154 & -0.153 & -0.358 & -0.356 \\
  15 & -0.080 & -0.082 & -0.311 & -0.371 & -0.158 & -0.157 & -0.373 & -0.371 \\
  16 & -0.067 & -0.070 & -0.326 & -0.386 & -0.156 & -0.155 & -0.389 & -0.386 \\
  17 & -0.069 & -0.072 & -0.370 & -0.424 & -0.172 & -0.170 & -0.428 & -0.424 \\
  18 & -0.060 & -0.064 & -0.415 & -0.426 & -0.167 & -0.166 & -0.430 & -0.426 \\
  19 & \textBF{-0.036} & \textBF{-0.037} & -0.404 & -0.435 & -0.167 & -0.164 & -0.442 & -0.435 \\
  20 & -0.058 & \textBF{-0.052} & -0.415 & -0.456 & -0.190 & -0.185 & -0.470 & -0.456 \\\hline
  Mean & -0.055 & -0.056 & -0.234 & -0.274 & -0.111 & -0.110 & -0.276 & -0.274 \\
  Median & -0.061 & -0.064 & -0.250 & -0.294 & -0.132 & -0.131 & -0.295 & -0.294 \\
\hline
$h$ & \multicolumn{4}{c|}{Optimal combination (OLS)} & \multicolumn{4}{c}{Optimal combination (GLS)} \\\hline
1   & 0.010 & 0.011 & \textBF{-0.021} & \textBF{-0.021} & \textBF{-0.002} & \textBF{-0.002} & -0.025 & -0.025 \\ 
2   & 0.011 & 0.011 & \textBF{-0.031} & \textBF{-0.031} & \textBF{-0.004} & \textBF{-0.004} & -0.035 & -0.035 \\ 
3   & \textBF{0.003} & \textBF{0.003} & \textBF{-0.050} & \textBF{-0.050} & -0.015 & -0.015 & -0.053 & -0.053 \\ 
4  & \textBF{-0.002} & \textBF{-0.002} & \textBF{-0.063} & \textBF{-0.063} & -0.021 & -0.021 & -0.066 & -0.066 \\ 
5   & \textBF{-0.007} & \textBF{-0.007} & \textBF{-0.078} & \textBF{-0.078} & -0.029 & -0.029 & -0.081 & -0.081 \\ 
6  & \textBF{-0.018} & \textBF{-0.018} & \textBF{-0.099} & \textBF{-0.100} & -0.042 & -0.042 & -0.102 & -0.102 \\ 
7   & \textBF{-0.027} & \textBF{-0.027} & \textBF{-0.116} & \textBF{-0.117} & -0.053 & -0.053 & -0.118 & -0.119 \\ 
8   & \textBF{-0.041} & \textBF{-0.041} & \textBF{-0.141} & \textBF{-0.141} & -0.068 & -0.068 & -0.142 & -0.142 \\ 
9   & \textBF{-0.051} & \textBF{-0.050} & \textBF{-0.160} & \textBF{-0.160} & -0.079 & -0.079 & \textBF{-0.160} & \textBF{-0.160} \\ 
10   &\textBF{-0.063} & \textBF{-0.063} & -0.182 & -0.181 & -0.093 & -0.092 & \textBF{-0.181} & \textBF{-0.180} \\ 
11   & \textBF{-0.068} & \textBF{-0.068} & -0.196 & -0.194 & -0.099 & -0.098 & \textBF{-0.194} & \textBF{-0.193} \\ 
12   & \textBF{-0.068} & \textBF{-0.068} & -0.203 & -0.202 & -0.100 & -0.099 & \textBF{-0.200} & \textBF{-0.199} \\ 
13   & \textBF{-0.067} & \textBF{-0.067} & -0.211 & -0.210 & -0.100 & -0.100 & \textBF{-0.207} & \textBF{-0.206} \\ 
14   & \textBF{-0.066} & \textBF{-0.066} & -0.218 & -0.216 & -0.101 & -0.100 & \textBF{-0.211} & \textBF{-0.210} \\ 
15   & \textBF{-0.063} & \textBF{-0.063} & -0.224 & -0.223 & -0.099 & -0.098 & \textBF{-0.215} & \textBF{-0.215} \\ 
16  & \textBF{-0.052} & \textBF{-0.051} & -0.228 & -0.226 & -0.089 & -0.088 & \textBF{-0.216} & \textBF{-0.215} \\ 
17   & \textBF{-0.057} & \textBF{-0.057} & -0.254 & -0.252 & -0.096 & -0.095 & \textBF{-0.238} & \textBF{-0.236} \\ 
18   & \textBF{-0.050} & \textBF{-0.051} & -0.247 & -0.248 & -0.086 & -0.086 & \textBF{-0.231} & \textBF{-0.230} \\ 
19   & -0.043 & -0.044 & -0.240 & -0.241 & -0.076 & -0.076 & \textBF{-0.220} & \textBF{-0.220} \\ 
20   & \textBF{-0.052} & \textBF{-0.052} & -0.253 & -0.248 & -0.090 & -0.088 & \textBF{-0.234} & \textBF{-0.228} \\ \hline
Mean   & \textBF{-0.039} & \textBF{-0.038} & -0.161 & -0.160 & -0.067 & -0.067 & \textBF{-0.156} & \textBF{-0.156} \\ 
Median   & \textBF{-0.050} & \textBF{-0.051} & -0.189 & -0.188 & -0.083 & -0.083 & \textBF{-0.188} & \textBF{-0.187} \\ \bottomrule
\end{tabular}
\end{small}
\caption{One-step-ahead to 20-step-ahead MFE $(\times 100)$ comparison between the different forecasting methods applied to the Australian infant mortality rates. For clarity of presentation, the MFEs have been multiplied by 100, in order to keep two decimal places. The bold entries highlight the method that has the smallest bias for each level of the hierarchy and each forecast horizon.}
\label{tab:AUS_MFE}
\end{table} 

\begin{table}[!htbp]
\centering
\tabcolsep 0.12in
\begin{small}
\begin{tabular}{@{}lrrrr|rrrr@{}}\toprule \\[-3.45ex] 
& Total & Sex & Region & Sex$\times$ Region & Total & Sex & Region & Sex$\times$ Region \\
$h$  & \multicolumn{4}{c|}{Base} & \multicolumn{4}{c}{Bottom-up}  \\
  \hline
  1 & 0.037 & 0.039 & 0.097 & 0.140 & 0.040 & 0.041 & 0.118 & 0.140 \\
  2 & \textBF{0.040} & \textBF{0.040} & 0.104 & 0.152 & \textBF{0.040} & 0.041 & 0.128 & 0.152 \\
  3 & \textBF{0.043} & \textBF{0.045} & 0.120 & 0.168 & 0.050 & 0.052 & 0.148 & 0.168 \\
  4 & \textBF{0.056} & 0.059 & 0.132 & 0.182 & 0.063 & 0.063 & 0.163 & 0.182 \\
  5 & \textBF{0.064} & \textBF{0.065} & 0.153 & 0.202 & 0.073 & 0.073 & 0.186 & 0.202 \\
  6 & 0.078 & 0.078 & 0.179 & 0.226 & 0.083 & 0.083 & 0.212 & 0.226 \\
  7 & 0.076 & 0.079 & 0.192 & 0.239 & 0.091 & 0.091 & 0.225 & 0.239 \\
  8 & 0.084 & 0.085 & 0.214 & 0.260 & 0.106 & 0.106 & 0.251 & 0.260 \\
  9 & 0.085 & 0.087 & 0.231 & 0.278 & 0.113 & 0.114 & 0.269 & 0.278 \\
  10 & 0.089 & 0.091 & 0.251 & 0.299 & 0.127 & 0.127 & 0.295 & 0.299 \\
  11 & 0.090 & 0.091 & 0.264 & 0.315 & 0.136 & 0.136 & 0.312 & 0.315 \\
  12 & 0.090 & 0.091 & 0.274 & 0.329 & 0.141 & 0.141 & 0.325 & 0.329 \\
  13 & 0.089 & 0.090 & 0.289 & 0.349 & 0.147 & 0.146 & 0.345 & 0.349 \\
  14 & 0.085 & 0.087 & 0.301 & 0.366 & 0.154 & 0.153 & 0.363 & 0.366 \\
  15 & 0.080 & 0.082 & 0.316 & 0.381 & 0.158 & 0.157 & 0.379 & 0.381 \\
  16 & 0.067 & 0.070 & 0.331 & 0.396 & 0.156 & 0.155 & 0.392 & 0.396 \\
  17 & 0.069 & 0.072 & 0.370 & 0.434 & 0.172 & 0.170 & 0.430 & 0.434 \\
  18 & 0.060 & 0.066 & 0.418 & 0.445 & 0.167 & 0.166 & 0.441 & 0.445 \\
  19 & \textBF{0.036} & \textBF{0.037} & 0.411 & 0.465 & 0.167 & 0.164 & 0.463 & 0.465 \\
  20 & 0.058 & 0.052 & 0.432 & 0.495 & 0.190 & 0.185 & 0.499 & 0.495 \\\hline
  Mean & 0.069 & 0.070 & 0.254 & 0.306 & 0.119 & 0.118 & 0.297 & 0.306 \\
  Median & 0.073 & 0.075 & 0.257 & 0.307 & 0.132 & 0.131 & 0.304 & 0.307 \\
   \hline
$h$ & \multicolumn{4}{c|}{Optimal combination (OLS)} & \multicolumn{4}{c}{Optimal combination (GLS)} \\\hline
  1 & \textBF{0.032} & \textBF{0.037} & 0.093 & 0.124 & 0.036 & \textBF{0.037} & \textBF{0.090} & \textBF{0.119} \\ 
  2 & 0.043 & 0.045 & 0.103 & 0.132 & \textBF{0.040} & 0.041 & \textBF{0.093} & \textBF{0.125} \\ 
  3 & 0.047 & 0.048 & 0.120 & 0.144 & 0.045 & 0.046 & \textBF{0.109} & \textBF{0.135} \\ 
  4 & 0.058 & \textBF{0.059} & 0.130 & 0.153 & 0.057 & \textBF{0.059} & \textBF{0.115} & \textBF{0.140} \\ 
  5 & 0.068 & 0.067 & 0.145 & 0.168 & 0.065 & \textBF{0.065} & \textBF{0.131} & \textBF{0.154} \\ 
  6 & \textBF{0.072} & \textBF{0.071} & 0.166 & 0.184 & 0.076 & 0.075 & \textBF{0.151} & \textBF{0.170} \\ 
  7 & \textBF{0.067} & \textBF{0.068} & 0.174 & 0.188 & 0.076 & 0.076 & \textBF{0.156} & \textBF{0.173} \\ 
  8 & \textBF{0.069} & \textBF{0.071} & 0.185 & 0.200 & 0.084 & 0.084 & \textBF{0.172} & \textBF{0.186} \\ 
  9 & \textBF{0.067} & \textBF{0.071} & 0.194 & 0.208 & 0.088 & 0.088 & \textBF{0.182} & \textBF{0.195} \\ 
  10 & \textBF{0.068} & \textBF{0.069} & 0.206 & 0.217 & 0.094 & 0.094 & \textBF{0.196} & \textBF{0.204} \\ 
  11 & \textBF{0.068} & \textBF{0.068} & 0.215 & 0.223 & 0.099 & 0.098 & \textBF{0.203} & \textBF{0.210} \\ 
  12 & \textBF{0.068} & \textBF{0.068} & 0.220 & 0.232 & 0.100 & 0.099 & \textBF{0.207} & \textBF{0.218} \\ 
  13 & \textBF{0.067} & \textBF{0.067} & 0.235 & 0.246 & 0.100 & 0.100 & \textBF{0.217} & \textBF{0.226} \\ 
  14 & \textBF{0.066} & \textBF{0.066} & 0.241 & 0.256 & 0.101 & 0.100 & \textBF{0.221} & \textBF{0.232} \\ 
  15 & \textBF{0.063} & \textBF{0.063} & 0.260 & 0.270 & 0.099 & 0.098 & \textBF{0.226} & \textBF{0.236} \\ 
  16 & \textBF{0.052} & \textBF{0.054} & 0.271 & 0.281 & 0.089 & 0.088 & \textBF{0.226} & \textBF{0.233} \\ 
  17 & \textBF{0.057} & \textBF{0.057} & 0.302 & 0.315 & 0.096 & 0.095 & \textBF{0.250} & \textBF{0.259} \\ 
  18 & \textBF{0.050} & \textBF{0.051} & 0.325 & 0.331 & 0.086 & 0.086 & \textBF{0.261} & \textBF{0.267} \\ 
  19 & 0.043 & 0.044 & 0.332 & 0.339 & 0.076 & 0.076 & \textBF{0.258} & \textBF{0.271} \\ 
  20 & \textBF{0.052} & \textBF{0.052} & 0.344 & 0.345 & 0.090 & 0.088 & \textBF{0.278} & \textBF{0.284} \\ \hline
  Mean & \textBF{0.059} & \textBF{0.060} & 0.213 & 0.228 & 0.080 & 0.080 & \textBF{0.187} & \textBF{0.202} \\ 
  Median & \textBF{0.065} & \textBF{0.064} & 0.211 & 0.220 & 0.087 & 0.087 & \textBF{0.200} & \textBF{0.207} \\ 
   \bottomrule
\end{tabular}
\end{small}
\caption{One-step-ahead to 20-step-ahead MAFE $(\times 100)$ comparison between the different forecasting methods applied to the Australian infant mortality rates. For clarity of presentation, the MAFEs have been multiplied by 100, in order to keep two decimal places. The bold entries highlight the method that has the smallest forecast errors for each level of the hierarchy and each forecast horizon.}
\label{tab:AUS_MAFE}
\end{table}

\begin{table}[!htbp] 
\centering 
  \tabcolsep 0.12in
  \begin{small}
\begin{tabular}{@{}lcccc|cccc@{}}\toprule \\[-3.45ex] 
& Total & Sex & Region & Sex$\times$ Region & Total & Sex & Region & Sex$\times$ Region \\
$h$  & \multicolumn{4}{c|}{Base} & \multicolumn{4}{c}{Bottom-up}  \\
  \hline
 1 & 0.050 & 0.052 & 0.159 & 0.254 & 0.050 & 0.052 & 0.210 & 0.254 \\
  2 & \textBF{0.049} & 0.052 & 0.179 & 0.270 & 0.049 & \textBF{0.051} & 0.227 & 0.270 \\
  3 & \textBF{0.056} & \textBF{0.059} & 0.213 & 0.304 & 0.062 & 0.064 & 0.269 & 0.304 \\
  4 & \textBF{0.067} & 0.071 & 0.251 & 0.326 & 0.074 & 0.075 & 0.293 & 0.326 \\
  5 & 0.075 & 0.078 & 0.274 & 0.359 & 0.085 & 0.087 & 0.331 & 0.359 \\
  6 & 0.087 & 0.089 & 0.323 & 0.400 & 0.099 & 0.100 & 0.374 & 0.400 \\
  7 & 0.089 & 0.093 & 0.335 & 0.422 & 0.107 & 0.108 & 0.395 & 0.422 \\
  8 & 0.099 & 0.103 & 0.364 & 0.469 & 0.123 & 0.124 & 0.439 & 0.469 \\
  9 & 0.098 & 0.103 & 0.404 & 0.506 & 0.130 & 0.131 & 0.475 & 0.506 \\
  10 & 0.103 & 0.109 & 0.445 & 0.543 & 0.142 & 0.144 & 0.517 & 0.543 \\
  11 & 0.100 & 0.105 & 0.483 & 0.585 & 0.148 & 0.149 & 0.555 & 0.585 \\
  12 & 0.101 & 0.106 & 0.498 & 0.610 & 0.153 & 0.155 & 0.575 & 0.610 \\
  13 & 0.096 & 0.103 & 0.529 & 0.643 & 0.154 & 0.155 & 0.610 & 0.643 \\
  14 & 0.088 & 0.093 & 0.557 & 0.688 & 0.160 & 0.160 & 0.649 & 0.688 \\
  15 & 0.089 & 0.095 & 0.598 & 0.706 & 0.164 & 0.164 & 0.682 & 0.706 \\
  16 & 0.071 & 0.080 & 0.644 & 0.745 & 0.157 & 0.157 & 0.727 & 0.745 \\
  17 & 0.074 & 0.083 & 0.716 & 0.813 & 0.173 & 0.174 & 0.806 & 0.813 \\
  18 & 0.071 & 0.082 & 0.831 & 0.844 & 0.168 & 0.169 & 0.835 & 0.844 \\
  19 & \textBF{0.038} & \textBF{0.040} & 0.843 & 0.869 & 0.167 & 0.167 & 0.859 & 0.869 \\
  20 & 0.058 & \textBF{0.052} & 0.888 & 0.901 & 0.190 & 0.186 & 0.902 & 0.901 \\\hline
  Mean & 0.078 & 0.082 & 0.477 & 0.563 & 0.128 & 0.129 & 0.536 & 0.563 \\
  Median & 0.081 & 0.086 & 0.464 & 0.564 & 0.145 & 0.146 & 0.536 & 0.564 \\
     \hline
$h$ & \multicolumn{4}{c|}{Optimal combination (OLS)} & \multicolumn{4}{c}{Optimal combination (GLS)} \\\hline
1  & 0.049 & 0.053 & 0.149 & 0.199 & \textBF{0.048} & \textBF{0.050} & \textBF{0.137} & \textBF{0.186} \\ 
 2 & 0.055 & 0.059 & 0.155 & 0.204 & 0.049 & 0.052 & \textBF{0.138} & \textBF{0.188} \\ 
 3 & 0.057 & 0.061 & 0.197 & 0.235 & 0.057 & 0.059 & \textBF{0.171} & \textBF{0.211} \\ 
 4 & \textBF{0.067} & 0.072 & 0.216 & 0.252 & 0.067 & \textBF{0.070} & \textBF{0.185} & \textBF{0.223} \\ 
 5 & \textBF{0.073} & \textBF{0.076} & 0.245 & 0.274 & 0.076 & 0.078 & \textBF{0.209} & \textBF{0.240} \\ 
 6 & \textBF{0.079} & \textBF{0.082} & 0.279 & 0.306 & 0.085 & 0.087 & \textBF{0.237} & \textBF{0.267} \\ 
 7 & \textBF{0.079} & \textBF{0.082} & 0.286 & 0.317 & 0.089 & 0.091 & \textBF{0.241} & \textBF{0.275} \\ 
 8 & \textBF{0.080} & \textBF{0.083} & 0.321 & 0.353 & 0.099 & 0.101 & \textBF{0.271} & \textBF{0.304} \\ 
 9 & \textBF{0.077} & \textBF{0.081} & 0.346 & 0.378 & 0.100 & 0.103 & \textBF{0.292} & \textBF{0.323} \\ 
 10 & \textBF{0.078} & \textBF{0.082} & 0.379 & 0.403 & 0.107 & 0.109 & \textBF{0.319} & \textBF{0.343} \\ 
 11 & \textBF{0.075} & \textBF{0.078} & 0.407 & 0.435 & 0.107 & 0.109 & \textBF{0.341} & \textBF{0.366} \\ 
 12 & \textBF{0.075} & \textBF{0.077} & 0.418 & 0.449 & 0.109 & 0.110 & \textBF{0.345} & \textBF{0.374} \\ 
 13 & \textBF{0.072} & \textBF{0.075} & 0.441 & 0.472 & 0.105 & 0.107 & \textBF{0.362} & \textBF{0.391} \\ 
 14 & \textBF{0.068} & \textBF{0.069} & 0.464 & 0.498 & 0.104 & 0.105 & \textBF{0.376} & \textBF{0.407} \\ 
 15 & \textBF{0.069} & \textBF{0.070} & 0.492 & 0.513 & 0.104 & 0.105 & \textBF{0.392} & \textBF{0.413} \\ 
 16 & \textBF{0.053} & \textBF{0.058} & 0.523 & 0.539 & 0.090 & 0.091 & \textBF{0.410} & \textBF{0.427} \\ 
 17 & \textBF{0.059} & \textBF{0.061} & 0.589 & 0.596 & 0.097 & 0.098 & \textBF{0.462} & \textBF{0.470} \\ 
 18 & \textBF{0.053} & \textBF{0.058} & 0.607 & 0.618 & 0.089 & 0.091 & \textBF{0.475} & \textBF{0.491} \\ 
 19 & 0.043 & 0.046 & 0.616 & 0.624 & 0.076 & 0.076 & \textBF{0.475} & \textBF{0.483} \\ 
 20  & \textBF{0.052} & 0.057 & 0.646 & 0.651 & 0.090 & 0.088 & \textBF{0.500} & \textBF{0.514} \\ \hline
 Mean & \textBF{0.066} & \textBF{0.069} & 0.389 & 0.416 & 0.087 & 0.089 & \textBF{0.317} & \textBF{0.345} \\ 
 Median & \textBF{0.069} & \textBF{0.071} & 0.393 & 0.419 & 0.090 & 0.091 & \textBF{0.330} & \textBF{0.354} \\ 
   \bottomrule
\end{tabular} 
  \caption{One-step-ahead to 20-step-ahead RMSFE ($\times 100$) comparison between the different forecasting methods applied to the Australian infant mortality rates. For clarity of presentation, the RMSFEs have been multiplied by 100, in order to keep two decimal places. The bold entries highlight the method that performs the best for each level of the hierarchy and each forecast horizon.}  \label{tab:AUS_RMSFE}
\end{small}
\end{table}

Based on the MFE$(h)$, the optimal combination methods generally outperform the base and bottom-up forecasting methods. In the top level and Level 1, the optimal combination (OLS) method has smaller forecast bias than the optimal combination (GLS) method at all horizons, with exceptions of $h=1$ and $h=2$. At Level 2 and the bottom level, the forecasts obtained from the optimal combination (OLS) method have smaller forecast bias than the optimal combination (GLS) method at the shorter forecast horizons from $h=1$ to $h=9$, but less so at the longer forecast horizons.

Based on the MAFE$(h)$ and RMSFE$(h)$, the optimal combination methods generally outperform the base and bottom-up forecasting methods. In the top level and Level 1, the optimal combination (OLS) method has smaller forecast errors than the optimal combination (GLS) method at the medium to long forecast horizons, but less so at the shorter forecast horizons. At Level 2 and the bottom level, the forecasts obtained from the optimal combination (GLS) method outperforms the optimal combination (OLS) method for every forecast horizon. Averaging across all levels of a hierarchy, the point forecasts obtained from the optimal combination (GLS) method are the most accurate in all methods investigated, and the method produces reconciled forecasts that obey a grouped structure.

\begin{figure}[!htbp]
\centering
\includegraphics[width=\textwidth]{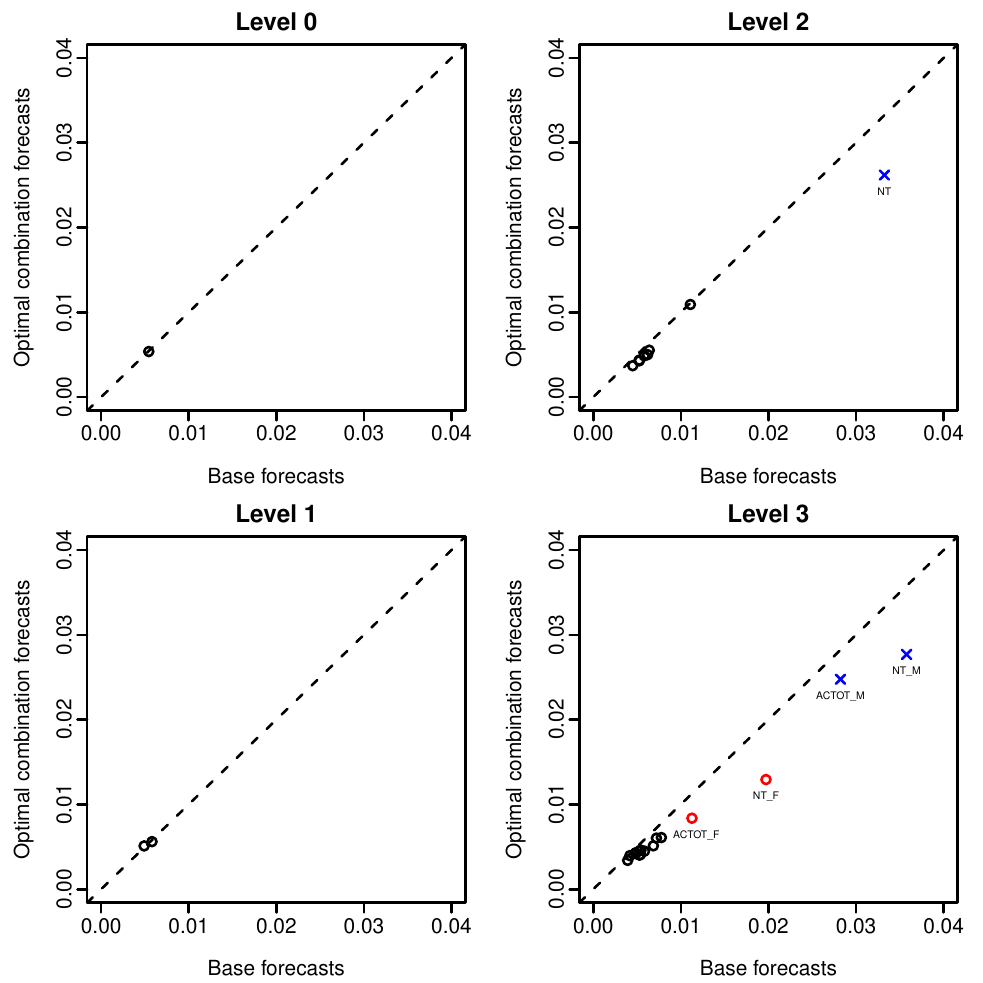}
\caption{Diagnostic plot of the 20-step-ahead forecasts at each level of the hierarchy for the Australian infant mortality between the base and optimal combination forecasts.}\label{fig:diag_1}
\end{figure}

To highlight the discrepancy in point forecasts between the base forecasts and optimal combination forecasts, we present a diagnostic plot showing the 20-step-ahead forecasts obtained from these two methods. As an illustration, since the base forecasts provide a foundation for the reconciled forecasts obtained from the optimal combination (OLS) method, the diagnostic plot allows us to visualize the forecasts that are similar or different between the two methods. As shown in Fig.~\ref{fig:diag_1}, there are almost no difference between the two methods at the top level and Level 1. At Level 2, there is only a difference for the NT region. At the bottom level, the largest differences for both sexes are ACTOT and NT regions.

\subsection{Influence of the $S$ matrix on point forecast accuracy}

The potential improvement in point forecast accuracy in the reconciliation methods relies crucially on the accurate forecast of the $S$ matrix. Recall that the $S$ matrix includes ratios of forecast exposure-at-risk. To forecast these exposure-at-risk, we again use the automatic ARIMA method to model and forecast exposure-at-risk at the logarithmic scale. By taking the exponential back-transformation, forecast exposure-at-risk in the original scale is obtained. In Tables~\ref{tab:MAFE_BU},~\ref{tab:MAFE_OLS} and~\ref{tab:MAFE_GLS}, we compare the MAFE among the reconciliation methods with forecast and holdout $S$ matrices. 
\begin{table}[!htbp]
\tabcolsep 0.36cm
\centering
\begin{tabular}{@{}lrrrrrrrr@{}}
\toprule
 $h$ & \multicolumn{2}{c}{Total} & \multicolumn{2}{c}{Sex} & \multicolumn{2}{c}{Region} & \multicolumn{2}{c}{Sex $\times$ Region}  \\ 
 & $\widehat{S}_{n+h|n}$ & $S_{n+h}$  & $\widehat{S}_{n+h|n}$ & $S_{n+h}$  & $\widehat{S}_{n+h|n}$ & $S_{n+h}$  & $\widehat{S}_{n+h|n}$ & $S_{n+h}$  \\
\midrule
  1 & \textBF{0.0399} & \textBF{0.0399} & 0.0406 & \textBF{0.0405} & 0.1175 & \textBF{0.1173} & \textBF{0.1397} & \textBF{0.1397} \\ 
  2 & \textBF{0.0404} & \textBF{0.0404} & 0.0407 & \textBF{0.0406} & \textBF{0.1275} & 0.1277 & \textBF{0.1516} & \textBF{0.1516} \\ 
  3 & 0.0504 & \textBF{0.0496} & 0.0515 & \textBF{0.0504} & \textBF{0.1482} & 0.1483 & \textBF{0.1681} & \textBF{0.1681} \\ 
  4 & 0.0629 & \textBF{0.0611} & 0.0627 & \textBF{0.0608} & \textBF{0.1625} & \textBF{0.1625} & \textBF{0.1816} & \textBF{0.1816} \\ 
  5 & 0.0728 & \textBF{0.0695} & 0.0727 & \textBF{0.0692} & 0.1862 & \textBF{0.1861} & \textBF{0.2022} & \textBF{0.2022} \\ 
  6 & 0.0831 & \textBF{0.0804} & 0.0831 & \textBF{0.0802} & \textBF{0.2120} & 0.2123 & \textBF{0.2264} & \textBF{0.2264} \\ 
  7 & 0.0908 & \textBF{0.0843} & 0.0907 & \textBF{0.0840} & \textBF{0.2253} & 0.2254 & \textBF{0.2387} & \textBF{0.2387} \\ 
  8 & 0.1059 & \textBF{0.0970} & 0.1058 & \textBF{0.0967} & \textBF{0.2511} & 0.2514 & \textBF{0.2598} & \textBF{0.2598} \\ 
  9 & 0.1126 & \textBF{0.1016} & 0.1145 & \textBF{0.1026} & \textBF{0.2693} & \textBF{0.2693} & \textBF{0.2782} & \textBF{0.2782} \\ 
  10 & 0.1270 & \textBF{0.1111} & 0.1266 & \textBF{0.1108} & 0.2953 & \textBF{0.2943} & \textBF{0.2990} & \textBF{0.2990} \\ 
  11 & 0.1364 & \textBF{0.1169} & 0.1360 & \textBF{0.1163} & 0.3125 & \textBF{0.3115} & \textBF{0.3145} & \textBF{0.3145} \\ 
  12 & 0.1414 & \textBF{0.1184} & 0.1411 & \textBF{0.1179} & 0.3252 & \textBF{0.3245} & \textBF{0.3289} & \textBF{0.3289} \\ 
  13 & 0.1467 & \textBF{0.1188} & 0.1464 & \textBF{0.1183} & 0.3454 & \textBF{0.3435} & \textBF{0.3487} & \textBF{0.3487} \\ 
  14 & 0.1536 & \textBF{0.1188} & 0.1527 & \textBF{0.1183} & 0.3627 & \textBF{0.3613} & \textBF{0.3661} & \textBF{0.3661} \\ 
  15 & 0.1576 & \textBF{0.1167} & 0.1566 & \textBF{0.1161} & 0.3787 & \textBF{0.3752} & \textBF{0.3806} & \textBF{0.3806} \\ 
  16 & 0.1560 & \textBF{0.1078} & 0.1547 & \textBF{0.1073} & 0.3915 & \textBF{0.3903} & \textBF{0.3961} & \textBF{0.3961}\\ 
  17 & 0.1720 & \textBF{0.1147} & 0.1703 & \textBF{0.1141} & 0.4302 & \textBF{0.4267} & \textBF{0.4341} & \textBF{0.4341} \\ 
  18 & 0.1673 & \textBF{0.1065} & 0.1659 & \textBF{0.1059} & 0.4413 & \textBF{0.4381} & \textBF{0.4453} & \textBF{0.4453}\\ 
  19 & 0.1670 & \textBF{0.0966} & 0.1643 & \textBF{0.0961} & 0.4632 & \textBF{0.4566} & \textBF{0.4651} & \textBF{0.4651}  \\ 
  20 & 0.1898 & \textBF{0.1222} & 0.1847 & \textBF{0.1218} & 0.4988 & \textBF{0.4861} & \textBF{0.4951} & \textBF{0.4951}  \\\midrule 
  Mean & 0.1187 & \textBF{0.0936} & 0.1181 & \textBF{0.0934} & 0.2972 & \textBF{0.2954} & \textBF{0.3060} & \textBF{0.3060} \\ 
  Median & 0.1317 & \textBF{0.1041} & 0.1313 & \textBF{0.1043} & 0.3039 & \textBF{0.3029} & \textBF{0.3067} & \textBF{0.3067}  \\ 
\bottomrule
\end{tabular}
\caption{A comparison of MAFE ($\times 100$) between the bottom-up method with forecast $S$ and actual $S$ matrices.}\label{tab:MAFE_BU}
\end{table}

\begin{table}[ht]
\tabcolsep 0.36cm
\centering
\begin{tabular}{@{}lrrrrrrrr@{}}
\toprule
 $h$ & \multicolumn{2}{c}{Total} & \multicolumn{2}{c}{Sex} & \multicolumn{2}{c}{Region} & \multicolumn{2}{c}{Sex$\times$Region} \\ 
 & $\widehat{S}_{n+h|n}$ & $S_{n+h}$  & $\widehat{S}_{n+h|n}$ & $S_{n+h}$  & $\widehat{S}_{n+h|n}$ & $S_{n+h}$  & $\widehat{S}_{n+h|n}$ & $S_{n+h}$   \\
\midrule
  1 & \textBF{0.0324} & \textBF{0.0324} & \textBF{0.0374} & 0.0377 & 0.0927 & \textBF{0.0926} & \textBF{0.1238} & \textBF{0.1238}  \\ 
  2 & \textBF{0.0434} & 0.0438 & \textBF{0.0452} & 0.0456 & \textBF{0.1026} & 0.1028 & \textBF{0.1322} & \textBF{0.1322} \\ 
  3 & \textBF{0.0466} & 0.0472 & \textBF{0.0484} & 0.0488 & 0.1198 & \textBF{0.1196} & \textBF{0.1442} & \textBF{0.1442}  \\ 
  4 & \textBF{0.0578} & 0.0579 & \textBF{0.0594} & 0.0595 & \textBF{0.1299} & \textBF{0.1299} & \textBF{0.1535} & \textBF{0.1535} \\ 
  5 & \textBF{0.0678} & \textBF{0.0678} & \textBF{0.0671} & 0.0673 & \textBF{0.1455} & \textBF{0.1455} & \textBF{0.1675} & \textBF{0.1675}  \\ 
  6 & 0.0717 & \textBF{0.0707} & 0.0713 & \textBF{0.0703} & \textBF{0.1656} & \textBF{0.1656} & \textBF{0.1842} & \textBF{0.1842} \\ 
  7 & 0.0672 & \textBF{0.0658} & 0.0675 & \textBF{0.0658} & \textBF{0.1742} & 0.1747 & \textBF{0.1884} & \textBF{0.1884} \\ 
  8 & 0.0695 & \textBF{0.0644} & 0.0706 & \textBF{0.0672} & \textBF{0.1849} & \textBF{0.1849} & \textBF{0.1998} & \textBF{0.1998} \\ 
  9 & 0.0674 & \textBF{0.0606} & 0.0709 & \textBF{0.0630} & 0.1945 & \textBF{0.1940} & \textBF{0.2080} & \textBF{0.2080} \\ 
  10 & 0.0681 & \textBF{0.0556} & 0.0686 & \textBF{0.0574} & 0.2065 & \textBF{0.2056} & \textBF{0.2167} & \textBF{0.2167} \\ 
  11 & 0.0683 & \textBF{0.0520} & 0.0685 & \textBF{0.0536} & 0.2149 & \textBF{0.2144} & \textBF{0.2232} & \textBF{0.2232} \\ 
  12 & 0.0682 & \textBF{0.0493} & 0.0678 & \textBF{0.0489} & 0.2204 & \textBF{0.2202} & \textBF{0.2324} & \textBF{0.2324} \\ 
  13 & 0.0672 & \textBF{0.0446} & 0.0669 & \textBF{0.0465} & 0.2354 & \textBF{0.2347} & \textBF{0.2463} & \textBF{0.2463} \\ 
  14 & 0.0662 & \textBF{0.0384} & 0.0656 & \textBF{0.0381} & 0.2412 & \textBF{0.2403} & \textBF{0.2555} & \textBF{0.2555} \\ 
  15 & 0.0632 & \textBF{0.0397} & 0.0626 & \textBF{0.0424} & 0.2596 & \textBF{0.2581} & \textBF{0.2700} & \textBF{0.2700} \\ 
  16 & 0.0516 & \textBF{0.0173} & 0.0542 & \textBF{0.0269} & 0.2707 & \textBF{0.2701} & \textBF{0.2814} & \textBF{0.2814} \\ 
  17 & 0.0573 & \textBF{0.0192} & 0.0569 & \textBF{0.0222} & 0.3019 & \textBF{0.3010} & \textBF{0.3151} & \textBF{0.3151} \\ 
  18 & 0.0499 & \textBF{0.0205} & 0.0508 & \textBF{0.0331} & 0.3253 & \textBF{0.3226} & \textBF{0.3309} & \textBF{0.3309} \\ 
  19 & 0.0429 & \textBF{0.0123} & 0.0442 & \textBF{0.0161} & 0.3325 & \textBF{0.3308} & \textBF{0.3391} & \textBF{0.3391} \\ 
  20 & 0.0518 & \textBF{0.0025} & 0.0515 & \textBF{0.0294} & 0.3438 & \textBF{0.3366} & \textBF{0.3446} & \textBF{0.3446}  \\ \midrule
  Mean & 0.0589 & \textBF{0.0431} & 0.0598 & \textBF{0.0470} & 0.2131 & \textBF{0.2122} & \textBF{0.2278} & \textBF{0.2278} \\ 
  Median & 0.0647 & \textBF{0.0459} & 0.0641 & \textBF{0.0477} & 0.2107 & \textBF{0.2100} & \textBF{0.2200} & \textBF{0.2200}  \\ 
\bottomrule
\end{tabular}
\caption{A comparison of MAFE $(\times 100)$ between the optimal combination method (the OLS estimator) with forecast $S$ and actual $S$ matrices.}\label{tab:MAFE_OLS}
\end{table}

\begin{table}[!htbp]
\tabcolsep 0.36cm
\centering
\begin{tabular}{@{}lrrrrrrrr@{}}
\toprule
 $h$ & \multicolumn{2}{c}{Total} & \multicolumn{2}{c}{Sex} & \multicolumn{2}{c}{Region} & \multicolumn{2}{c}{Sex $\times$ Region}  \\ 
 & $\widehat{S}_{n+h|n}$ & $S_{n+h}$  & $\widehat{S}_{n+h|n}$ & $S_{n+h}$  & $\widehat{S}_{n+h|n}$ & $S_{n+h}$  & $\widehat{S}_{n+h|n}$ & $S_{n+h}$  \\
\midrule
  1 & 0.0363 & \textBF{0.0362} & \textBF{0.0366} & 0.0367 & \textBF{0.0897} & \textBF{0.0897} & \textBF{0.1189} & \textBF{0.1189} \\ 
  2 & \textBF{0.0404} & 0.0406 & \textBF{0.0407} & 0.0410 & \textBF{0.0935} & 0.0936 & \textBF{0.1247} & \textBF{0.1247} \\ 
  3 & \textBF{0.0450} & \textBF{0.0450} & 0.0462 & \textBF{0.0461} & 0.1089 & \textBF{0.1087} & \textBF{0.1354} & \textBF{0.1354} \\ 
  4 & \textBF{0.0567} & 0.0569 & \textBF{0.0587} & \textBF{0.0587} & \textBF{0.1151} & \textBF{0.1151} & \textBF{0.1400} & \textBF{0.1400} \\ 
  5 & \textBF{0.0646} & 0.0649 & 0.0651 & \textBF{0.0650} & 0.1313 & \textBF{0.1311} & \textBF{0.1538} & \textBF{0.1538} \\ 
  6 & 0.0756 & \textBF{0.0739} & 0.0752 & \textBF{0.0738} & \textBF{0.1512} & \textBF{0.1512} & \textBF{0.1705} & \textBF{0.1705} \\ 
  7 & 0.0761 & \textBF{0.0737} & 0.0765 & \textBF{0.0737} & \textBF{0.1561} & 0.1569 & \textBF{0.1733} & \textBF{0.1733} \\ 
  8 & 0.0840 & \textBF{0.0788} & 0.0839 & \textBF{0.0796} & \textBF{0.1725} & 0.1726 & \textBF{0.1864} & \textBF{0.1864} \\ 
  9 & 0.0879 & \textBF{0.0811} & 0.0880 & \textBF{0.0808} & 0.1822 & \textBF{0.1821} & \textBF{0.1948} & \textBF{0.1948} \\ 
  10 & 0.0939 & \textBF{0.0849} & 0.0941 & \textBF{0.0855} & 0.1961 & \textBF{0.1953} & \textBF{0.2043} & \textBF{0.2043} \\ 
  11 & 0.0986 & \textBF{0.0871} & 0.0981 & \textBF{0.0866} & 0.2032 & \textBF{0.2028} & \textBF{0.2101} & \textBF{0.2101} \\ 
  12 & 0.0995 & \textBF{0.0864} & 0.0992 & \textBF{0.0860} & \textBF{0.2069} & 0.2071 & \textBF{0.2178} & \textBF{0.2178} \\ 
  13 & 0.1004 & \textBF{0.0846} & 0.1001 & \textBF{0.0842} & 0.2166 & \textBF{0.2151} & \textBF{0.2258} & \textBF{0.2258} \\ 
  14 & 0.1007 & \textBF{0.0810} & 0.0999 & \textBF{0.0805} & 0.2210 & \textBF{0.2204} & \textBF{0.2318} & \textBF{0.2318} \\ 
  15 & 0.0990 & \textBF{0.0757} & 0.0982 & \textBF{0.0753} & 0.2262 & \textBF{0.2241} & \textBF{0.2357} & \textBF{0.2357} \\ 
  16 & 0.0888 & \textBF{0.0623} & 0.0880 & \textBF{0.0620} & 0.2264 & \textBF{0.2258} & \textBF{0.2332} & \textBF{0.2332} \\ 
  17 & 0.0957 & \textBF{0.0648} & 0.0947 & \textBF{0.0645} & 0.2498 & \textBF{0.2486} & \textBF{0.2590} & \textBF{0.2590} \\ 
  18 & 0.0865 & \textBF{0.0546} & 0.0864 & \textBF{0.0550} & 0.2614 & \textBF{0.2593} & \textBF{0.2667} & \textBF{0.2667} \\ 
  19 & 0.0764 & \textBF{0.0395} & 0.0760 & \textBF{0.0394} & 0.2580 & \textBF{0.2553} & \textBF{0.2709} & \textBF{0.2709} \\ 
  20 & 0.0899 & \textBF{0.0594} & 0.0875 & \textBF{0.0596} & 0.2776 & \textBF{0.2707} & \textBF{0.2837} & \textBF{0.2837} \\ \midrule
  Mean & 0.0798 & \textBF{0.0666} & 0.0797 & \textBF{0.0667} & 0.1872 & \textBF{0.1863} & \textBF{0.2018} & \textBF{0.2018} \\ 
  Median & 0.0872 & \textBF{0.0693} & 0.0869 & \textBF{0.0693} & 0.1996 & \textBF{0.1991} & \textBF{0.2072} & \textBF{0.2072}  \\ 
\bottomrule
\end{tabular}
\caption{A comparison of MAFE $(\times 100)$ between the optimal combination method (the GLS estimator) with forecast $S$ and actual $S$ matrices.}\label{tab:MAFE_GLS}
\end{table}

At the top two levels, more accurate point forecasts can be obtained by using the holdout $S$ matrix. At the bottom two levels, there are comparably smaller differences in point forecast accuracy between the forecast and actual $S$ matrices.

\section{Results of the interval forecasts}\label{sec:5}

As described in Section~\ref{sec:2.4}, we constructed pointwise prediction intervals using the maximum entropy and parametric bootstrap methods. The maximum entropy bootstrap method generates bootstrap samples that preserve the correlation in the original time series, whereas the parametric bootstrap method generates bootstrap forecasts for each bootstrap sample. Based on these bootstrap forecasts, we assess the variability of point forecasts by constructing prediction intervals based on quantiles. By averaging over all bootstrap prediction intervals, we obtain the averaged prediction intervals. For a reasonably large level of significance $\alpha$, such as $\alpha=0.2$, averaging prediction intervals works well as we estimate the center distribution of the quantiles. Due to heavy computational cost, there are 100 bootstrap samples obtained by a maximum entropy bootstrap. Within each bootstrap sample, the number of parametric bootstrap forecasts is 100.

\begin{figure}[!htbp]
\centering
\includegraphics[width=\textwidth]{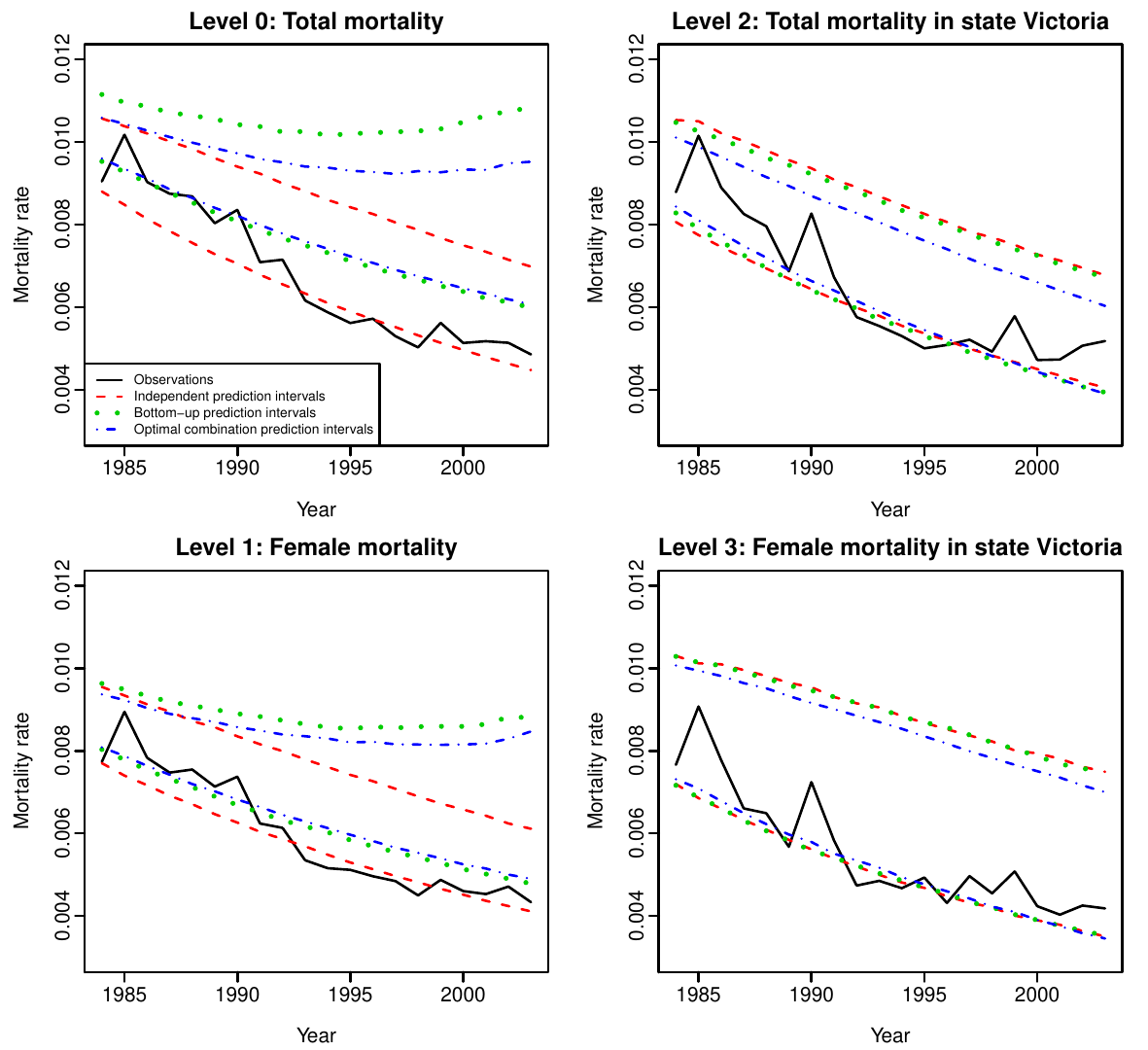}
\caption{Based on the Australian infant mortality data from 1933 to 1983, we produce 20-steps-ahead prediction intervals for years 1984 to 2003, at the nominal coverage probability of 80\%. For ease of presentation, we show the 80\% prediction intervals for a few selected series at each level of the hierarchy.}\label{fig:interval}
\end{figure}

Figure~\ref{fig:interval} shows the 80\% pointwise averaged prediction intervals of the direct 20-steps-ahead Australian infant mortality rate forecasts for a few selected series at each level of the hierarchy from 1984 to 2003. At the top level, there seems to be a larger difference in interval forecasts between the base forecasting and two grouped time-series methods. From the middle to bottom levels, the interval forecasts are very similar between the three methods. For the optimal combination (GLS) method, the construction of prediction interval is hindered by the difficulty in measuring forecast uncertainty associated with $\bm{\Sigma}^+$, and thus we leave this for future research.

\subsection{Interval forecast evaluation}

At a nominal level of 80\%, prediction intervals are constructed by taking corresponding quantiles, where the lower bound is denoted by $\widehat{L}_{\omega+h,j}$ and the upper bound is denoted by $\widehat{U}_{\omega+h,j}$ for $j=1,\dots,m$ and $m$ representing the total number of series in a hierarchy. With a pointwise prediction interval and its corresponding holdout data point in the forecasting period, we can assess interval forecast accuracy by the interval score of \cite{GR07}, defined as
\begin{small}
\begin{align*}
S_{\alpha,j}\left(\widehat{L}_{\omega+h,j}, \widehat{U}_{\omega+h,j}, Y_{\omega+h,j}\right) = \left(\widehat{U}_{\omega+h,j} - \widehat{L}_{\omega+h,j}\right) &+ \frac{2}{\alpha}\left(\widehat{L}_{\omega+h,j} - Y_{\omega+h,j}\right) \mathds{1}\left\{Y_{\omega+h,j} < \widehat{L}_{\omega+h,j}\right\} \\
&+\frac{2}{\alpha}\left(Y_{\omega+h,j} - \widehat{U}_{\omega+h,j}\right)\mathds{1}\left\{Y_{\omega+h,j}>\widehat{U}_{\omega+h,j}\right\},
\end{align*}
\end{small}
\hspace{-.13in} where $Y_{\omega+h,j}$ represents the holdout samples in the forecasting period for the series $j$, and $\mathds{1}\{\cdot\}$ is a binary indicator function. This interval score combines the halfwidth of the prediction intervals with the coverage probability difference between the nominal and empirical coverage probabilities. Intuitively, a forecaster is rewarded for narrow prediction intervals, but a penalty is incurred, the size of which depends on the level of significance $\alpha$, if the holdout samples lie outside the prediction intervals.

For each series $j$ at each forecast horizon, we obtain
\begin{equation*}
\overline{S}_{\alpha,j}(h) = \frac{1}{(21-h)}\sum^{n+(20-h)}_{\omega=n}S_{\alpha,j}\left(\widehat{L}_{\omega+h,j}, \widehat{U}_{\omega+h,j}, Y_{\omega+h,j}\right), \qquad h=1,\dots,20, 
\end{equation*}
where $S_{\alpha,j}\left(\widehat{L}_{\omega+h,j}, \widehat{U}_{\omega+h,j}, Y_{\omega+h,j}\right)$ denotes the interval score at each level of the hierarchy for the holdout samples in the forecasting period. By averaging the interval score $\overline{S}_{\alpha,j}(h)$ across the number of series within each level of a hierarchy, we obtain an overall assessment of the interval forecast accuracy for each level within a hierarchy. The mean interval score is then defined by
\begin{equation*}
\overline{S}_{\alpha,k}(h) = \frac{1}{m_k}\sum^{m_k}_{j=1}\overline{S}_{\alpha,j}(h),
\end{equation*}
where $m_k$ denotes the number of series at the $k^{\text{th}}$ level of the hierarchy, for $k=1,\dots,K$.

\subsection{Interval forecast accuracy of Australian infant mortality}

In Table~\ref{tab:interval_score}, we present the mean interval scores for the one-step-ahead to 20-step-ahead forecasts at each level of the hierarchy between the three methods. For ease of comparison, we highlight in bold the method that performs the best for each level of the hierarchy and each forecast horizon, based on the smallest $\overline{S}_{\alpha, k}(h)$.

\begin{table}[!htbp] 
\centering 
\tabcolsep 0.095in
\begin{small}
\begin{tabular}{@{\extracolsep{5pt}} lcccc|cccc@{}} 
\toprule \\[-3.45ex] 
& Total & Sex & Region & Sex$\times$ Region & Total & Sex & Region & Sex$\times$ Region \\
$h$  & \multicolumn{4}{c|}{Base} & \multicolumn{4}{c}{Bottom-up} \\
\hline
1 & $\textBF{0.17}$ & $\textBF{0.18}$ & $\textBF{0.55}$ & $0.77$ & $0.24$ & $0.22$ & $0.63$ & $0.76$ \\
2 & $\textBF{0.17}$ & $\textBF{0.21}$ & $\textBF{0.59}$ & $0.81$ & $0.27$ & $0.24$ & $0.70$ & $0.81$ \\
3 & $\textBF{0.21}$ & $\textBF{0.26}$ & $\textBF{0.65}$ & $0.88$ & $0.36$ & $0.31$ & $0.78$ & $0.88$ \\
4 & $\textBF{0.23}$ & $\textBF{0.28}$ & $\textBF{0.69}$ & $0.91$ & $0.46$ & $0.38$ & $0.82$ & $0.91$ \\
5 & $\textBF{0.26}$ & $\textBF{0.32}$ & $\textBF{0.74}$ & $0.96$ & $0.54$ & $0.45$ & $0.89$ & $0.96$ \\
6 & $\textBF{0.31}$ & $\textBF{0.35}$ & $\textBF{0.80}$ & $1.02$ & $0.65$ & $0.53$ & $1.01$ & $1.02$ \\
7 & $\textBF{0.34}$ & $\textBF{0.37}$ & $\textBF{0.81}$ & $1.08$ & $0.73$ & $0.60$ & $1.10$ & $1.08$ \\
8 & $\textBF{0.37}$ & $\textBF{0.39}$ & $\textBF{0.87}$ & $1.18$ & $0.85$ & $0.68$ & $1.22$ & $1.18$ \\
9 & $\textBF{0.35}$ & $\textBF{0.38}$ & $\textBF{0.93}$ & $1.28$ & $0.94$ & $0.74$ & $1.32$ & $1.28$ \\
10 & $\textBF{0.37}$ & $\textBF{0.40}$ & $\textBF{1.02}$ & $1.36$ & $1.05$ & $0.85$ & $1.43$ & $1.36$ \\
11 & $\textBF{0.35}$ & $\textBF{0.38}$ & $\textBF{1.07}$ & $1.47$ & $1.12$ & $0.90$ & $1.52$ & $1.49$ \\
12 & $\textBF{0.36}$ & $\textBF{0.38}$ & $\textBF{1.07}$ & $1.52$ & $1.16$ & $0.92$ & $1.58$ & $1.52$ \\
13 & $\textBF{0.29}$ & $\textBF{0.34}$ & $\textBF{1.13}$ & $1.62$ & $1.20$ & $0.95$ & $1.69$ & $1.62$ \\
14 & $\textBF{0.26}$ & $\textBF{0.29}$ & $\textBF{1.16}$ & $1.74$ & $1.25$ & $0.99$ & $1.80$ & $1.73$ \\
15 & $\textBF{0.30}$ & $\textBF{0.31}$ & $\textBF{1.17}$ & $1.84$ & $1.30$ & $1.02$ & $1.97$ & $1.82$ \\
16 & $\textBF{0.24}$ & $\textBF{0.24}$ & $\textBF{1.26}$ & $1.89$ & $1.27$ & $0.99$ & $2.12$ & $1.88$ \\
17 & $\textBF{0.24}$ & $\textBF{0.24}$ & $\textBF{1.46}$ & $2.05$ & $1.40$ & $1.11$ & $2.33$ & $2.06$ \\
18 & $\textBF{0.25}$ & $\textBF{0.25}$ & $\textBF{1.78}$ & $2.19$ & $1.39$ & $1.09$ & $2.48$ & $2.18$ \\
19 & $\textBF{0.24}$ & $\textBF{0.24}$ & $\textBF{1.96}$ & $\textBF{2.19}$ & $1.42$ & $1.11$ & $2.46$ & $2.20$ \\
20 & $\textBF{0.25}$ & $\textBF{0.24}$ & $\textBF{2.19}$ & $2.36$ & $1.62$ & $1.29$ & $2.59$ & $\textBF{2.28}$  \\\hline 
Mean & $\textBF{0.28}$ & $\textBF{0.30}$ & $\textBF{1.10}$ & $1.46$ & $0.96$ & $0.77$ & $1.52$ & $1.45$ \\
Median & $\textBF{0.26}$ & $\textBF{0.30}$ & $\textBF{1.04}$ & $1.42$ & $1.09$ & $0.87$ & $1.48$ & $1.42$ \\  
\hline
$h$ & \multicolumn{4}{c|}{Optimal combination (OLS)}  \\\cline{1-5}
1 & $0.27$ & $0.21$ & $0.60$ & $\textBF{0.72}$ \\ 
2 & $0.28$ & $0.23$ & $0.68$ & $\textBF{0.76}$ \\ 
3 & $0.37$ & $0.30$ & $0.76$ & $\textBF{0.83}$ \\  
4 & $0.48$ & $0.37$ & $0.82$ & $\textBF{0.86}$ \\
5 & $0.55$ & $0.44$ & $0.91$ & $\textBF{0.92}$ \\ 
6 & $0.66$ & $0.54$ & $1.05$ & $\textBF{0.97}$ \\ 
7 & $0.74$ & $0.59$ & $1.15$ & $\textBF{1.02}$ \\  
8 & $0.87$ & $0.69$ & $1.27$ & $\textBF{1.11}$ \\ 
9 & $0.95$ & $0.75$ & $1.37$ & $\textBF{1.20}$ \\ 
10 & $1.06$ & $0.85$ & $1.49$ & $\textBF{1.27}$ \\
11 & $1.13$ & $0.89$ & $1.59$ & $\textBF{1.36}$ \\ 
12 & $1.16$ & $0.90$ & $1.64$ & $\textBF{1.41}$ \\ 
13 & $1.19$ & $0.92$ & $1.73$ & $\textBF{1.50}$ \\ 
14 & $1.22$ & $0.95$ & $1.83$ & $\textBF{1.58}$ \\ 
15 & $1.24$ & $0.96$ & $2.00$ & $\textBF{1.67}$ \\
16 & $1.21$ & $0.91$ & $2.16$ & $\textBF{1.71}$ \\ 
17 & $1.31$ & $1.01$ & $2.44$ & $\textBF{1.88}$ \\  
18 & $1.31$ & $1.00$ & $2.66$ & $\textBF{2.09}$ \\
19 & $1.34$ & $1.01$ & $2.78$ & $\textBF{2.19}$ \\ 
20 & $1.56$ & $1.22$ & $2.99$ & $2.40$ \\\cline{1-5}
Mean & $0.94$ & $0.74$ & $1.59$ & $\textBF{1.37}$ \\ 
Median & $1.09$ & $0.87$ & $1.54$ & $\textBF{1.32}$ \\ 
\bottomrule \\[-1.8ex] 
\end{tabular} 
\end{small}
\caption{One-step-ahead to 20-step-ahead interval score $(\times 100)$ comparison between the different forecasting methods applied to the Australian infant mortality rates. Note that the slight discrepancy between the base forecasts and bottom-up forecasts at the bottom level is due to different random seeds used in bootstrapping.}\label{tab:interval_score}
\end{table}

Based on the overall interval forecast accuracy $\overline{S}_{\alpha,k}(h)$, the base forecasting method gives the most accurate interval forecasts at the top three levels, but the optimal combination method demonstrates the best interval forecast accuracy for the bottom-level series. Averaged over all levels of the hierarchy, the base forecasting method outperforms the two grouped time-series methods in terms of mean interval scores. A possible explanation for the inferior interval accuracy of the grouped time-series forecasting methods is that they require the accurate forecasts of the $S$ matrix consisting of the forecast exposure-to-risk, which may introduce additional forecast uncertainty. However, from a viewpoint of forecast interpretation, the grouped time series methods produce interval forecasts that obey a grouped time-series structure.

Due to the limited space, although not shown in the paper, the grouped time-series forecasting methods can improve interval forecast accuracy in another data set, namely the Japanese data set \citep{JMD15}. When the forecasts of the exposure-to-risk are accurate, the reconciliation of interval mortality forecasts are more accurate than the base interval forecasts. These results can be obtained upon request from the author.

\subsection{Influence of the $S$ matrix on interval forecast accuracy}

The potential improvement in interval forecast accuracy in the reconciliation methods relies crucially on the accurate forecast of the $S$ matrix. Recall that the $S$ matrix includes ratios of forecast exposure-at-risk. To obtain bootstrap forecasts of these exposure-at-risk, we use the parametric bootstrap and maximum entropy bootstrap methods to simulate future samples of the exposure-at-risk at the logarithmic scale. By taking the exponential back-transformation, bootstrap forecasts of exposure-at-risk in the original scale are obtained. In Tables~\ref{tab:interval_score_BU} and~\ref{tab:interval_score_optim}, we compare the interval score among the reconciliation methods with forecast and holdout $S$ matrices.

\begin{table}[!htbp]
\tabcolsep 0.44cm
\centering
\begin{tabular}{@{}lrrrrrrrr@{}}
  \toprule
  $h$ & \multicolumn{2}{c}{Total} & \multicolumn{2}{c}{Sex} & \multicolumn{2}{c}{Region} & \multicolumn{2}{c}{Sex $\times$ Region} \\ 
   & $\widehat{S}_{n+h|n}$ & $S_{n+h}$  & $\widehat{S}_{n+h|n}$ & $S_{n+h}$  & $\widehat{S}_{n+h|n}$ & $S_{n+h}$  & $\widehat{S}_{n+h|n}$ & $S_{n+h}$ \\
  \midrule
  1 & 0.24 & \textBF{0.23} & \textBF{0.22} & \textBF{0.22} & \textBF{0.63} & \textBF{0.63} & \textBF{0.76} & \textBF{0.76} \\ 
  2 & 0.27 & \textBF{0.25} & 0.24 & \textBF{0.23} & \textBF{0.70} & \textBF{0.70} & \textBF{0.81} & \textBF{0.81} \\ 
  3 & 0.36 & \textBF{0.33} & 0.31 & \textBF{0.30} & \textBF{0.78} & \textBF{0.78} & \textBF{0.88} & \textBF{0.88} \\ 
  4 & 0.46 & \textBF{0.42} & 0.38 & \textBF{0.37} & \textBF{0.82} & 0.83 & \textBF{0.91} & \textBF{0.91} \\ 
  5 & 0.54 & \textBF{0.49} & 0.45 & \textBF{0.43} & \textBF{0.89} & 0.92 & \textBF{0.96} & \textBF{0.96} \\ 
  6 & 0.65 & \textBF{0.59} & 0.53 & \textBF{0.50} & \textBF{1.01} & 1.06 & \textBF{1.02} & \textBF{1.02} \\ 
  7 & 0.73 & \textBF{0.64} & 0.60 & \textBF{0.55} & \textBF{1.10} & 1.16 & \textBF{1.08} & \textBF{1.08} \\ 
  8 & 0.85 & \textBF{0.76} & 0.69 & \textBF{0.63} & \textBF{1.22} & 1.29 & 1.18 & \textBF{1.17} \\ 
  9 & 0.94 & \textBF{0.81} & 0.74 & \textBF{0.67} & \textBF{1.32} & 1.39 & \textBF{1.28} & \textBF{1.28} \\ 
  10 & 1.05 & \textBF{0.90} & 0.85 & \textBF{0.75} & \textBF{1.43} & 1.52 & 1.36 & \textBF{1.35} \\ 
  11 & 1.12 & \textBF{0.93} & 0.90 & \textBF{0.77} & \textBF{1.52} & 1.63 & 1.49 & \textBF{1.47} \\ 
  12 & 1.16 & \textBF{0.94} & 0.92 & \textBF{0.76} & \textBF{1.58} & 1.69 & \textBF{1.52} & 1.53 \\ 
  13 & 1.20 & \textBF{0.94} & 0.95 & \textBF{0.75} & \textBF{1.69} & 1.83 & \textBF{1.62} & \textBF{1.62} \\ 
  14 & 1.25 & \textBF{0.94} & 0.99 & \textBF{0.75} & \textBF{1.80} & 1.95 & \textBF{1.73} & 1.74 \\ 
  15 & 1.30 & \textBF{0.92} & 1.02 & \textBF{0.72} & \textBF{1.97} & 2.14 & \textBF{1.82} & \textBF{1.82} \\ 
  16 & 1.27 & \textBF{0.83} & 0.99 & \textBF{0.62} & \textBF{2.12} & 2.29 & 1.88 & \textBF{1.86} \\ 
  17 & 1.40 & \textBF{0.88} & 1.11 & \textBF{0.67} & \textBF{2.33} & 2.52 & 2.06 & \textBF{2.04} \\ 
  18 & 1.39 & \textBF{0.83} & 1.09 & \textBF{0.64} & \textBF{2.48} & 2.68 & 2.18 & \textBF{2.17} \\ 
  19 & 1.42 & \textBF{0.77} & 1.11 & \textBF{0.56} & \textBF{2.46} & 2.70 & 2.20 & \textBF{2.17} \\ 
  20 & 1.62 & \textBF{0.97} & 1.29 & \textBF{0.75} & \textBF{2.59} & 2.83 & \textBF{2.28} & 2.32 \\ \midrule
  Mean & 0.96 & \textBF{0.72} & 0.77 & \textBF{0.58} & \textBF{1.52} & 1.63 & \textBF{1.45} & \textBF{1.45} \\ 
  Median & 1.09 & \textBF{0.82} & 0.87 & \textBF{0.63} & \textBF{1.48} & 1.57 & 1.42 & \textBF{1.41} \\ 
\bottomrule
\end{tabular}
\caption{A comparison of interval score $(\times 100)$ between the bottom-up method with forecast $S$ and actual $S$ matrices.}\label{tab:interval_score_BU}
\end{table}

\begin{table}[!htbp]
\tabcolsep 0.44cm
\centering
\begin{tabular}{@{}lrrrrrrrr@{}}
  \toprule
  $h$ & \multicolumn{2}{c}{Total} & \multicolumn{2}{c}{Sex} & \multicolumn{2}{c}{Region} & \multicolumn{2}{c}{Sex $\times$ Region} \\ 
   & $\widehat{S}_{n+h|n}$ & $S_{n+h}$  & $\widehat{S}_{n+h|n}$ & $S_{n+h}$  & $\widehat{S}_{n+h|n}$ & $S_{n+h}$  & $\widehat{S}_{n+h|n}$ & $S_{n+h}$ \\\midrule
  1 & \textBF{0.27} & \textBF{0.27} & \textBF{0.21} & 0.23 & \textBF{0.60} & 0.61 & \textBF{0.72} & 0.73 \\ 
  2 & 0.28 & \textBF{0.27} & \textBF{0.23} & \textBF{0.23} & \textBF{0.68} & 0.69 & \textBF{0.76} & \textBF{0.76} \\ 
  3 & 0.37 & \textBF{0.35} & \textBF{0.30} & \textBF{0.30} & \textBF{0.76} & 0.78 & \textBF{0.83} & 0.84 \\ 
  4 & 0.48 & \textBF{0.46} & \textBF{0.37} & \textBF{0.37} & \textBF{0.82} & 0.85 & \textBF{0.86} & 0.87 \\ 
  5 & 0.55 & \textBF{0.51} & 0.44 & \textBF{0.43} & \textBF{0.91} & 0.94 & \textBF{0.92} & \textBF{0.92} \\ 
  6 & 0.66 & \textBF{0.61} & 0.54 & \textBF{0.52} & \textBF{1.05} & 1.09 & \textBF{0.97} & 0.98 \\ 
  7 & 0.74 & \textBF{0.67} & 0.59 & \textBF{0.56} & \textBF{1.15} & 1.19 & \textBF{1.02} & 1.03 \\ 
  8 & 0.87 & \textBF{0.79} & 0.69 & \textBF{0.64} & \textBF{1.27} & 1.33 & \textBF{1.11} & 1.12 \\ 
  9 & 0.95 & \textBF{0.85} & 0.75 & \textBF{0.69} & \textBF{1.37} & 1.44 & \textBF{1.20} & 1.21 \\ 
  10 & 1.06 & \textBF{0.94} & 0.85 & \textBF{0.77} & \textBF{1.49} & 1.58 & \textBF{1.27} & 1.28 \\ 
  11 & 1.13 & \textBF{0.98} & 0.89 & \textBF{0.79} & \textBF{1.59} & 1.67 & \textBF{1.36} & \textBF{1.36} \\ 
  12 & 1.16 & \textBF{0.98} & 0.90 & \textBF{0.78} & \textBF{1.64} & 1.75 & \textBF{1.41} & \textBF{1.41} \\ 
  13 & 1.19 & \textBF{0.98} & 0.92 & \textBF{0.77} & \textBF{1.73} & 1.88 & \textBF{1.50} & \textBF{1.50} \\ 
  14 & 1.22 & \textBF{0.97} & 0.95 & \textBF{0.76} & \textBF{1.83} & 1.99 & \textBF{1.58} & \textBF{1.58} \\ 
  15 & 1.24 & \textBF{0.94} & 0.96 & \textBF{0.73} & \textBF{2.00} & 2.19 & \textBF{1.67} & \textBF{1.67} \\ 
  16 & 1.21 & \textBF{0.85} & 0.91 & \textBF{0.63} & \textBF{2.16} & 2.34 & 1.71 & \textBF{1.69} \\ 
  17 & 1.31 & \textBF{0.90} & 1.01 & \textBF{0.67} & \textBF{2.44} & 2.62 & 1.88 & \textBF{1.87} \\ 
  18 & 1.31 & \textBF{0.84} & 1.00 & \textBF{0.64} & \textBF{2.66} & 2.84 & 2.09 & \textBF{2.02} \\ 
  19 & 1.34 & \textBF{0.77} & 1.01 & \textBF{0.55} & \textBF{2.78} & 2.94 & 2.19 & \textBF{2.16} \\ 
  20 & 1.56 & \textBF{0.98} & 1.22 & \textBF{0.74} & \textBF{2.99} & 3.17 & 2.40 & \textBF{2.38} \\ \midrule
  Mean & 0.94 & \textBF{0.75} & 0.74 & \textBF{0.59} & \textBF{1.59} & 1.70 & \textBF{1.37} & \textBF{1.37} \\ 
  Median & 1.09 & \textBF{0.85} & 0.87 & \textBF{0.64} & \textBF{1.54} & 1.63 & \textBF{1.32} & \textBF{1.32} \\ 
  \bottomrule
\end{tabular}
\caption{A comparison of interval score $(\times 100)$ between the optimal combination method with forecast $S$ and actual $S$ matrices.}
\label{tab:interval_score_optim}
\end{table}

At the top two levels, more accurate interval forecasts can be obtained by using the holdout $S$ matrix. At the Region level, the forecast $S$ matrix gives a smaller interval score than the holdout $S$ matrix. This rather surprising result may due to the forecast uncertainty associated with the mortality rates. At the bottom level, there is no difference in terms of interval forecast accuracy between the forecast and actual $S$ matrices.

\section{Conclusions}\label{sec:6}

This article adapts a bottom-up method and an optimal combination method for modeling and forecasting grouped time series of infant mortality rates. The bottom-up method models and forecasts time series at the bottom level and then aggregates to the top level using the summing matrix. The optimal combination method optimally combines the base forecasts through linear regression by generating a set of revised forecasts that are as close as possible to the base forecasts but that also aggregate consistently within the group. Under a mild assumption, regression coefficient can be estimated by either OLS or GLS estimator.

Using the regional infant mortality rates in Australia, we implemented these two grouped time-series forecasting methods that reconcile forecasts across different levels of a hierarchy. Furthermore, we compared the one-step-ahead to 20-step-ahead point forecast accuracy, and found that the optimal combination method has the smallest overall forecast error in the Australian data set considered. 

Through the maximum entropy and parametric bootstrap methods, we present a means of constructing pointwise prediction intervals for grouped time series. The maximum entropy bootstrap is capable of mimicking the correlation within and between the original time series. For each bootstrapped time series, we can then fit an optimal ARIMA model and generate forecasts; from these forecasts the corresponding prediction intervals are obtained. Averaging over all prediction intervals, we obtain averaged prediction intervals to evaluate forecast uncertainty associated with the point forecasts. 

In the Australian data set, we found that the base forecasting method gives the best overall interval forecast accuracy, but the two grouped time-series forecasting methods produce interval forecasts that obey a group structure and thus ease of interpretation. It is noteworthy that the accuracy of the reconciliation methods crucially depends on the forecast accuracy of the summing matrix. Although the forecast $S$ matrix does not differ much from the holdout $S$ matrix, the reconciliation methods enjoy improved forecast accuracy with the holdout $S$ matrix at the top and middle levels, but less so at the bottom level.

There are several ways in which this study could be further extended and we briefly outline five of these. First, the methods are proposed from a frequentist viewpoint, and they can be compared with a hierarchical Bayesian method. Secondly, the methodology can be applied to cause-specific mortality, considered in \cite{ML97}, \cite{GK08} and \cite{GS15}. Thirdly, the methodology can be applied to other demographic data, such as population size. Fourthly, forecasts can also be obtained by multivariate time-series forecasting methods, such as vector autoregressive models, in order to take into account possible correlations between and within multiple time series. Finally, the idea of grouped time series can be extended to functional time series \citep[see][]{SH16}, where each series is a time series of functions, such as age-specific demographic rates. This work provides a natural foundation for such extensions.

%\section*{Acknowledgments}

%The author thanks two referees for their insightful suggestions and comments, which led to a much improved manuscript. The author is also grateful for the comments and suggestions received from Professors Rob Hyndman, Juha Alho, Peter W. F. Smith, Jakub Bijak, John Bryant and the participants at the $35^{\text{th}}$ International Symposium on Forecasting held at Riverside, California in June 2015. This research is funded by a Research School Faculty Grant from the Australian National University.

\newpage
\renewcommand\refname{REFERENCES}
\bibliographystyle{apacite}
\bibliography{hts}

\newpage

\begin{center}
\large Appendix: Maximum entropy bootstrap algorithm
\end{center}

An overview of the maximum entropy bootstrap algorithm is provided for generating a random realization of a univariate time series $x_t$. Consult \cite{Vinod04} for more details and an example. In the maximum entropy bootstrap algorithm,

\begin{enumerate}
\item Sort the original data in increasing order to create order statistics $x_{(t)}$ and store the ordering index vector.
\item Compute intermediate points $z_t = \frac{x_{(t)}+x_{(t+1)}}{2}$ for $t=1,\dots,n-1$ from the order statistics.
\item Compute the trimmed mean, denoted by $m_{\text{trim}}$ of deviations $x_t - x_{t-1}$ among our consecutive observations. Compute the lower limit for the left tail as $z_0 = x_{(1)} - m_{\text{trim}}$ and the upper limit for the right tail as $z_n = x_{(n)}+m_{\text{trim}}$. These limits become the limiting intermediate points.
\item Compute the mean of the maximum entropy density within each interval such that the ``mean-preserving constraint" is satisfied. Interval means are denoted as $m_t$. The means for the first and last intervals have simpler formulas:
\[ \left\{ \begin{array}{l}
         m_1=0.75 x_{(1)}+0.25 x_{(2)} \\
        m_k = 0.25x_{(k-1)} + 0.5 x_{(k)} + 0.25 x_{(k+1)}, \qquad k=2,\dots,n \\
        m_n = 0.25 x_{(n-1)} + 0.75 x_{(n)}
        \end{array} \right. \] 
\item Generate random numbers from Uniform$[0,1]$, compute sample quantiles of the maximum entropy density at those points and sort them.
\item Re-order the sorted sample quantiles by using the ordering index of Step 1. This recovers the time dependence relationships of the originally observed data.
\item Repeat Steps 2 to 6 several times.
\end{enumerate}

\end{document}